\documentclass[prd,aps,twocolumn,floatfix]{revtex4}

\usepackage{amssymb,graphicx}
\usepackage{epsfig}
\usepackage[usenames]{color}

\newcommand{\secapproach}{II}

\def\rmd{{\rm d}}

\newcommand{\had}{{\sc had}}

\begin{document}

\title{Understanding possible electromagnetic counterparts to loud gravitational wave events: 
Binary black hole effects on electromagnetic fields}

\author{Carlos Palenzuela${}^{1,2}$, Luis Lehner${}^{3,4,5}$ and Shin Yoshida${}^6$}

\affiliation{
${}^1$Canadian Institute for Theoretical Astrophysics (CITA), Toronto, Canada \\
${}^2$Max-Planck-Institut f\" ur Gravitationsphysik,
Albert-Einstein-Institut, Golm, Germany\\
${}^3$Perimeter Institute for Theoretical Physics, Waterloo, Ontario N2L 2Y5, Canada\\
${}^4$Department of Physics, University of Guelph,
Guelph, Ontario N1G 2W1, Canada, \\
${}^5$ Canadian Institute For Advanced Research (CIFAR),  Cosmology and Gravity Program, Canada\\
${}^6$Department of Physics and Astronomy, University of Tokyo,
Tokyo, Japan
}

\date{\today}

%
%
\begin{abstract}
In addition to producing loud gravitational waves (GW), the dynamics
of a binary black hole system could induce emission of electromagnetic (EM) radiation
by affecting the behavior of plasmas and electromagnetic fields
in their vicinity. We here study how the electromagnetic fields are affected 
by a pair of orbiting black holes through the merger. 
In particular, we show how the binary's dynamics induce
a  variability in possible electromagnetically induced emissions as well as an
enhancement of electromagnetic fields during the late-merge and merger epochs. 
These time dependent features will likely leave their imprint in processes
generating detectable emissions and can be exploited in the
detection of electromagnetic counterparts of gravitational waves. 
\end{abstract}

\maketitle

%
%
\section{Introduction}
The promise of detecting and analysing compact systems with both gravitational and electromagnetic
waves stands out as one of the most exciting prospects in the coming decades. As already pointed out
in a number of works (e.g.~\cite{Sylvestre:2003vc,Stubbs:2007mk}), most astrophysical systems which produce 
strong gravitational waves likely emit copiously in the electromagnetic band. Indeed the strong, and possibly highly
dynamical, gravitational fields around compact objects affect the dynamics of plasmas and matter which in turn induce
different emission mechanisms.

One example of such  a system is a black hole  surrounded by an accretion disk.
Strong emission from these systems is understood as the
result of radiative processes within jets powered by the extraction of rotational 
and binding energy. While the latter is qualitatively understood in terms of Newtonian-rooted
arguments based on the potential of the central object, the former relies
on extracting energy from a rotating black hole in the most efficient energy convertion process we know of.\\
\indent
The pioneering works of Penrose~\cite{Penrose:1969pc} and Blandford and
Znajek~\cite{Blandford:1977ds}, together with  a large body of subsequent work, has provided a basic understanding of
possible mechanisms to explain highly energetic emissions from single black hole systems
interacting with surrounding plasmas (e.g.~\cite{2000PhR...325...83L}). The interaction of electromagnetic
field lines with the strong gravitational field of a rotating black hole is the fundamental component of these mechanisms
to explain the  acceleration of
particles that traverse the black hole's ergosphere. This scenario of a
pseudo-stationary, single black hole  interacting with an accretion disk is reasonably well understood, and it is
employed to explain energetic phenomena such as gamma ray bursts, AGNs, quasars, blazars, etc. However,
a highly dynamical stage may occur prior to such a  pseudo-stationary regime which could
give rise to strong emissions. In the context of galaxy mergers, such a stage would naturally occur as individual
black holes in each galaxy eventually collide in the galaxy resulting from the merger~\cite{1980Natur.287..307B}.\\
\indent
Gravitational waves from such collision would be detectable by the Laser Interferometric Space Antenna (LISA) and,
as pointed out in e.g.~\cite{Milosavljevic:2004cg,Haiman:2009te}, the late orbitting and merger phases would take place
within a circumbinary disk and possibly even interacting with some residual plasma inside the orbiting black holes \cite{Chang:2009rx}.

Possible emissions from these systems are only understood at late times, when the
the pseudo-stationary picture mentioned above is applicable. However the intermediate regime has only recently been
approached by a few works~\cite{Palenzuela:2009yr,Chang:2009rx,vanMeter:2009gu} and in all cases with important
simplifications introduced to track the system. In this work, as a follow up of~\cite{Palenzuela:2009yr}, we concentrate
on understanding the dynamics of possible electromagnetic fields anchored in the circumbinary disk in the presence
of the merging black holes. In particular we examine the field configuration,
possible energy enhancement and time
variability of these fields as the merger take place and point out possible process that could give rise
to a signal around the merger time. To this end, we consider
the Einstein-Maxwell system in a setup that describes a pair of black holes close to the
merger epoch, study the electromagnetic field behavior and compare to the single black hole case. While we do
not consider a plasma in our current study, our analysis helps to understand the possible behavior in its
presence and lay the foundations to future work in this direction.\\
This work is organized as follows, 
in Section~\secapproach, we briefly review our formulation and numerical implementation of the problem.  
Section III describes the physical set up to study single and binary black hole configurations. Section IV discusses the results obtained for both scenarios considered  
and highlight main features which could induce emission with particular patterns.
We conclude with section V which offers some final considerations and discussions.   
%
%
%
\section{Formulation and numerical approach}

We solve the coupled Einstein-Maxwell system to model the black hole merger
interacting with an externally sourced magnetic field. We here
describe the particular formulation of the d
equations employed in our simulations.  We begin with a brief review
the 3+1 decomposition og Genera Relativity followed by a discussion
of the Cauchy problem for both Einstein and Maxwell equations.

\subsection{The 3+1 decomposition}
In the Cauchy or 3+1 formulation, the spacetime $(M,g_{ab}$) ($a,b=0,1,2,3$) 
is foliated with spacelike 
hypersurfaces labeled by constant coordinate time $x^0\equiv t = {\rm const}$. The metric of
these hypersurfaces is $\gamma_{ij} = g_{ij}$ ($i,j=1,2,3$). The normal vector to the hypersurfaces is
$n_a \equiv - \nabla_a t / ||\nabla_a t ||$, and coordinates defined on 
neighboring hypersurfaces can be related through the lapse function, $\alpha$, 
and shift vector, $\beta^i$.  With these definitions, the spacetime
line element can be expressed as
\begin{eqnarray}
\rmd s^2 &=& g_{ab}\, \rmd x^a \rmd x^b \nonumber \\
         &=&-\alpha^2 \, \rmd t^2 
            + \gamma_{ij}\left(\rmd x^i + \beta^i\, \rmd t\right)
                         \left(\rmd x^j + \beta^j\, \rmd t\right)\, ;
\end{eqnarray}
while the normal vector/covector is given explicity by
\begin{eqnarray}
n^a = \frac{1}{\alpha} (1,-\beta^i) ~~,~~ n_a &=& (-\alpha, 0)~~.
\end{eqnarray}
Indices on spacetime quantities are raised and lowered with the 4-metric,
$g_{ab}$, and its inverse, while the 3-metric $\gamma_{ij}$ and its inverse
are used to raise and lower indices on spatial quantities.
The following simple expressions relate the $3+1$ basic 
variables \{$\gamma_{ij}$, $\alpha$, $\beta^i$\} with
the four-dimensional metric \{$g_{ab}$\} by
\begin{eqnarray}
  \gamma_{ij} &=& g_{ij}~~,~~\alpha = \sqrt{-1/g^{00}}~~,
        ~~\beta^i = \gamma^{ij} g_{0j}~~.
\end{eqnarray}
In what follows we will make use of both sets of variables.


\subsection{Einstein equations}
The Einstein equations in the Generalized Harmonic
 formulation~\cite{Frie85,Garfinkle:2001ni} (GH) can be written as a system of ten nonlinear partial
differential equations for the spacetime metric $g_{ab}$. 
\begin{eqnarray}\label{dedonder1}
 && g^{cd} \partial_{cd} ~g_{ab}
 + \partial_{a} H_{b} + \partial_{b} H_{a}  =
 - 16~\pi~\left(T_{ab} - \frac{T}{2}~g_{ab}\right) \nonumber \\
 && + 2~\Gamma_{cab}H^c
  + 2~g^{cd}g^{ef}\bigg(\partial_e g_{ac}~ \partial_f g_{bd}
 - \Gamma_{ace}~\Gamma_{bdf}\bigg).
\end{eqnarray}
where the coordinates $x^a$ can be chosen to satisfy the generalized harmonic condition
\begin{equation}
\label{eq:harmonic}
\nabla^c \nabla_c x^a = - g^{bc} {\Gamma^a}_{bc} = H^a \label{harmonic1} \, ,
\end{equation}
for some arbitrary functions $H_a$ .
One possibility for determining these functions, which we adopt here,  
employs the original harmonic condition $H_i =0$ (i.e. for the spatial components),
while a damped wave equation for the time one \cite{Pret05}:
\begin{eqnarray}
\label{eq:harmonic_frans}
\nabla^c \nabla_c H_t &=& - \xi_1 \frac{\alpha-1}{\alpha^n} + \xi_2 n^c \partial_c H_t \, .
\end{eqnarray}
Since there is no coupling between the principal part of the Einstein equations
(\ref{dedonder1}) with the generalized harmonic condition (\ref{eq:harmonic_frans}),
the full system of equations (Einstein equations + gauge condition) is trivially
hyperbolic, determined by wave-like equations with non-linear source terms.

A reduction to first order of the evolution system (ie, 
the Einstein equations (\ref{dedonder1}) with
the generalized harmonic condition (\ref{eq:harmonic_frans}) )
can be achieved by introducing new independent variables related
to the time and space derivatives of the fields
\begin{eqnarray}
 \quad Q_{ab} &\equiv& - n^c \, \partial_c g_{ab} \, ,
\quad  D_{iab} \equiv \partial_i g_{ab} \, , \\
 \quad G &\equiv& - n^c \, \partial_c H_{t} \, ,
\quad  G_{i} \equiv \partial_i H_{t} \, .
\label{eq:firstorder}
\end{eqnarray}
With these definitions we can write the evolution equations in our GH
formalism in the following way \cite{Lindblom:2005qh,Palenzuela:2006wp}
\begin{eqnarray}
  \label{EE_geq}
  \partial_t g_{ab} &=& \beta^k~D_{kab} - \alpha~Q_{ab}, \\
  \partial_t Q_{ab} &=& \beta^k~\partial_k Q_{ab}
  - \alpha \gamma^{ij} \partial_i D_{jab} \nonumber \\ 
  &-& \alpha~ \partial_a H_b - \alpha~ \partial_b H_a +
  2~\alpha~ \Gamma_{cab}~ H^c \nonumber \\
  &+& 2\, \alpha\, g^{cd}~(\gamma^{ij} D_{ica} D_{jdb} - Q_{ca} Q_{db}
                   - g^{ef} \Gamma_{ace} \Gamma_{bdf}) \nonumber \\
  &-& \frac{\alpha}{2} n^c n^d Q_{cd} Q_{ab}
  - \alpha~\gamma^{ij} D_{iab} Q_{jc} n^c \nonumber \\ 
  &-& 8 \pi \, \alpha(2T_{ab} - g_{ab} T) \nonumber \\
  &-& 2 \sigma_0 \, \alpha \, [n_a Z_b + n_b Z_a - g_{a b} n^c Z_c ]  \nonumber  \\
  &+& \sigma_1 \, \beta^i ( D_{iab} -  \partial_i g_{ab} ),  \\
   \label{EE_Deq}
   \partial_t D_{iab} &=& \beta^k \partial_k D_{iab}
  - \alpha~\partial_i Q_{ab} \nonumber \\ 
   &+& \frac{\alpha}{2} n^c n^d D_{icd} Q_{ab}
  + \alpha~\gamma^{jk} n^c D_{ijc} D_{kab} \nonumber \\ 
   &-& \sigma_1 \, \alpha \, ( D_{iab} - \partial_i g_{ab} ) , 
\nonumber \\
  \partial_t H_{t} &=& \beta^k~G_{k} - \alpha~G, \\
  \partial_t G &=& \beta^k~\partial_k G
  - \alpha \gamma^{ij} \partial_i G_{j} \nonumber \\ 
  &-& \frac{\alpha}{2} G\, n^c n^d Q_{cd}
  - \alpha~\gamma^{ij} G_{i} Q_{jc} n^c + \alpha \Gamma_c G^c
\nonumber \\ 
  &+& \sigma_1 \, \beta^i ( G_{i} -  \partial_i H_{t} )
 - \xi_1 \frac{\alpha-1}{\alpha^{n-1}} -\alpha \xi_2 G \, , \\
   \label{HH_Deq}
   \partial_t G_{i} &=& \beta^k \partial_k G_{i}
  - \alpha~\partial_i G \nonumber \\ 
   &+& \frac{\alpha}{2} G n^c n^d D_{icd}
  + \alpha~\gamma^{jk} G_{k} n^c D_{ijc} \nonumber \\ 
   &-& \sigma_1 \, \alpha \, ( G_{i} - \partial_i H_{t} ) .
  \label{HH_geqend}
\end{eqnarray}
This GH formulation includes a number of constraints that must be satisfied 
for consistency. Namely, two sets of first order constraints ${\cal C}_{iab},
{\cal C}_{ijab}$ and the four-vector $Z_a$ accounting for the physical
energy and momentum constraints~\cite{2003PhRvD..67j4005B,Lindblom:2005qh},
\begin{eqnarray}\label{constraints}
  {\cal C}_{iab} &\equiv & \partial_i g_{ab} - D_{iab} = 0 ~,~~\nonumber \\
  {\cal C}_{ijab} &\equiv & \partial_i D_{jab} - \partial_j D_{iab} = 0 ~\nonumber \\
   2 Z_a &\equiv& -\Gamma_a - H_a(t,x^i) = 0 \, .
\end{eqnarray}
These constraints are controlled dynamically via the inclusion of a constraint
damping mechanism~\cite{Gundlach:2005eh}, by adding certain terms proportional to these constraints 
(with free parameters $\sigma_0$ and $\sigma_1$) to the evolution equations.


\subsection{Maxwell equations}
The Maxwell equations can be written in covariant form as
\begin{eqnarray}
  \nabla_{b}\, F^{a b} &=& 4 \pi I^{\,a}
\label{Maxwell_eqs1a} , \\
  \nabla_{b}\, {}^*F^{a b} &=& 0
\label{Maxwell_eqs1b}
\end{eqnarray}
where $F^{a b}$ is the Maxwell tensor of the electromagnetic
field, ${}^*F^{a b}$ is the Faraday tensor and $I^{a}$ is
the electric current 4-vector. Since $F^{a b}$ is
antisymmetric, the divergence of equation
(\ref{Maxwell_eqs1a}) leads to the current conservation
\begin{eqnarray}
   \nabla_{a} I^{\,a} = 0 ~.
\label{conserved_current}
\end{eqnarray}

When both the electric and magnetic susceptibility of the medium
vanish, as in vacuum or in a highly ionized plasma, the Faraday
tensor is simply the dual of the Maxwell one, that is
\begin{equation}
  {}^*F^{a b} = \frac{1}{2}\, \epsilon^{a b c d} \, F_{cd}
\label{Fduals}
\end{equation}
where $\epsilon^{a b c d}$ is the Levi-Civita
pseudotensor of the spacetime, which can be written in terms of
the 4-indices Levi-Civita symbol $\eta^{a b c d}$ as
\begin{eqnarray}\label{levicivita}
   \epsilon^{a b c d} = \frac{1}{\sqrt{g}}~
   \eta^{a b c d} \qquad
    \epsilon_{a b c d} = -\sqrt{g}~
   \eta^{a b c d}~.
\end{eqnarray}\index{Levi-Civita pseudotensor}
For intuitive and practical reasons, it is convenient to introduce and work 
with the electric and magnetic fields which are defined as
\begin{equation}
E^a = F^{ab} n_b ~~ , ~~  B^a = {}^*F^{a b} n_b ~~ \, .
\end{equation}
The vectors $E^{a}$ and $B^{a}$ are
the electric and magnetic fields measured by the normal observer and
are purely spatial (ie, $E^{a} n_{a} = B^{a} n_{a} = 0$).
The Faraday tensor can then be re-expressed as,
\begin{eqnarray}\label{Faraday tensor}
  F^{a b} &=& n^{a} E^{b} - n^{b} E^{a}
               + \epsilon^{abcd}~B_{c}\,n_{d} \, ,
\label{F_em1a} \\
  ^*F^{a b} &=& n^{a} B^{b} - n^{b} B^{a}
               - \epsilon^{abcd}~E_{c}\,n_{d} \, .
\label{F_em1b}
\end{eqnarray}
We consider here an extended Maxwell system \cite{2007MNRAS.382..995K} defined as
\begin{eqnarray}
  \nabla_{a} (F^{a b} + g^{a b} \Psi) &=& - 4 \pi I^{b}
  - \sigma_2\, n^{b} \Psi \, ,
\label{Maxwell_ext_eqs2a} \\
\nabla_{a} (^*F^{a b} + g^{a b} \phi) &=& -\sigma_2\, n^{b} \phi \, ,
\label{Maxwell_ext_eqs2b}
\end{eqnarray}
which reduces to the standard Maxwell equations if
$\Psi=0=\phi$. These  extra scalar fields play the
role ``error fields'' as they are tightly coupled to constraints
violations. Moreover, their induced evolution equations, obtained by considering
the divergence of the extended Maxwell equations, imply
\begin{eqnarray}
  \nabla_{a} \nabla^{a} \Psi &=& 
      - \nabla_{a} \left( \sigma_2\, n^{a} \Psi \right)
\label{Maxwell_sub1a} \\
\nabla_{a} \nabla^{a} \phi &=& 
      -  \nabla_{a} \left( \sigma_2\, n^{a} \phi \right) \, ,
\label{Maxwell_sub1b}
\end{eqnarray}
which are a generalization of the telegraph equation. Their structure 
ensures the constraints will propagate at the speed of light and will be damped 
within a timescale given by $\sigma_2^{-1}$. This strategy is
similar to that defined in~\cite{Dedner2002} for ideal MHD case.

The 3+1 version of equations (\ref{Maxwell_ext_eqs2a}-\ref{Maxwell_ext_eqs2b}), which
are the ones implemented in our code, are
\begin{eqnarray}
  (\partial_t - {\cal L}_{\beta})\, E^{i}  &-&
  \epsilon^{ijk}\nabla_j (\,\alpha B_k\,)
   + \alpha\,\gamma^{ij} \nabla_j\,\Psi = \nonumber \\
   && \alpha\, trK\, E^i - 4 \pi \alpha J^{\,i} \, ,
\label{maxwellext_3+1_eq1a} \\
  (\partial_t - {\cal L}_{\beta})\, B^{i} &+&
  \epsilon^{ijk} \nabla_j (\,\alpha E_k\,)
   + \alpha\, \gamma^{ij} \nabla_j\, \phi =\nonumber \\
   && \alpha\, trK\, B^i \, ,
\label{maxwellext_3+1_eq1c} \\
  (\partial_t - {\cal L}_{\beta})\,\Psi &+& \alpha\, \nabla_i E^i =
   4 \pi \alpha\, q -\alpha \sigma_2\, \Psi \, ,
\label{maxwellext_3+1_eq1b} \\
  (\partial_t - {\cal L}_{\beta})\,\phi &+& \alpha\, \nabla_i B^i =
   -\alpha \sigma_2\, \phi
\label{maxwellext_3+1_eq1d}\,.
\end{eqnarray}
with $trK$ the trace of the extrinsic curvature and 
where we have decomposed the current four-vector $I^{a} = 
q n^a + J^a$, (with the current $J^a$ satisfying $J^a n_a = 0$).
Obviously the standard Maxwell equations in a curved background
are recovered for $\Psi=\phi=0$. From equations (\ref{maxwellext_3+1_eq1b}) 
and (\ref{maxwellext_3+1_eq1d}) it follows that $\Psi$ and $\phi$ can
be regarded as the normal-time integrals  of the standard
divergence constraints
\begin{equation}\label{Maxwell_divs}
    \nabla_i E^i =  4 \pi q \,\, , \,\, \qquad \nabla_i B^i = 0 \, .
\end{equation}

Finally, just reming that the sources are coupled to the geometry by means
of the stress energy tensor $T_{ab}$ and its trace
$T\equiv g^{ab} T_{ab}$. For our case of interest, the stress energy tensor
is given by,
\begin{eqnarray}\label{stress-em}
   T_{a b} = \frac{1}{4 \pi} \left[ {F_{a}}^{c} ~ F_{b c}
     - \frac{1}{2}\, g_{a b} ~ F^{c d} F_{c d}  \right]~~,
\end{eqnarray}
which depends quadratically on the electric and magnetic fields.


\subsection{Implementation}

Our code implements both systems of equations where the constraints, as mentioned, are kept under control via 
different but related damping mechanisms: constraint damping for the Einstein equations~\cite{Gundlach:2005eh} 
and the extended divergence cleaning for the Maxwell equations as explained in the previous section.
We adopt boundary conditions defined via a
combination of Sommerfeld and constraint preserving boundary conditions~\cite{RLS07} for both systems. 
To this end, a characteristic decomposition of the Generalized Harmonic formalism eqns. (\ref{EE_geq}-
\ref{HH_geqend}) is performed (at each hypersurface) with respect to the wave front 
propagation direction, given by a normalized spatial vector \textbf{m}. This vector is
orthogonal to a given boundary and belongs to and ordered orthonormal triad
$\{\textbf{l},\textbf{p},\textbf{m}\}$. This decomposition gives,
\begin{eqnarray}
  g_{ab}~,H_t ~~~~~~~ v&=&0\\
  D_{l ab} ~, D_{p ab}~,G_{l} ~, G_{p} ~~~~~~~  v&=&-\beta^m \\
  L^{\pm}_{ab} \equiv Q_{ab} - \sigma_2 g_{ab} \pm D_{mab} ~~~~~~~ v&=&-\beta^m \pm \alpha \\
  L^{\pm} \equiv G - \sigma_2 H_t \pm G_{m} ~~~~~~~ v&=&-\beta^m \pm \alpha
\label{GH_eigenvectors}
\end{eqnarray}
where the symbol $\{l,p,m\}$ replacing an index means the
projection along the corresponding vector. The boundary conditions
are applied only to the incoming modes (i.e., $\{ L^{-}_{ab}, L^{-}\}$
and $\{ D_{lab},D_{pab},G_{l},G_{p} \}$ if $\beta^m > 0$) through their
time derivatives. We here explain in detail the modes related to the metric modes,
while the modes related to $H_t$ are treated in an analogous way.
The Sommerfeld condition considered is of the type~\cite{RLS07}
\begin{equation}
   (\partial_t +\partial_r + \frac{1}{r})(g_{ab} - \eta_{ab}) = 0 ~~,
\label{Sommerfeld1}
\end{equation}
where $\eta_{ab}$ is just the Minkowski metric. By taking a time derivative
of this equation and rewriting it in terms of the incoming characteristic
fields, the final form of the Sommerfeld conditions results:
\begin{equation}
   \partial_t \left[ L^{-}_{ab} + (\sigma_2 - \frac{1}{r}) g_{ab} \right] = 0 ~~.
\label{Sommerfeld2}
\end{equation}
We apply boundary conditions for $\{ D_{lab},D_{pab} \}$ if $\beta^m > 0$.
In this case we use the constraint preserving boundary conditions already given
in \cite{Lindblom:2005qh}, where the original time derivatives are corrected
by the 4-index constraint defined in eqn. (\ref{constraints}), namely
\begin{eqnarray}
   \partial_t D_{lab} = \partial_t D_{lab} - \beta^m m^i P^{j}_l C_{ijab} ~~,~~\\
   \partial_t D_{pab} = \partial_t D_{pab} - \beta^m m^i P^{j}_p C_{ijab} ~~,~~
\label{CPBC1}
\end{eqnarray}
where $P_{ab} \equiv g_{ab} + n_a n_b - m_a m_b$ is the projection
tensor on the boundary surface.

The incoming modes of the electromagnetic fields are defined via 
maximally dissipative conditions on the time derivatives,
induced from the physical picture of a circumbinary disk present
beyond the computational domain. The complete set of
eigenvectors for the extended Maxwell equations
(\ref{maxwellext_3+1_eq1a}-\ref{maxwellext_3+1_eq1d}),
again with respect to ordered orthonormal triad
$\{\textbf{l},\textbf{p},\textbf{m}\}$, is given by the following
list of eigenfields propagating with light speed $-\beta^m \pm \alpha$:
\begin{eqnarray}
  E_l &\pm& B_p ~~,~~  E_p \mp B_l ~~,\\
  \Psi &\pm& E_m ~~,~~ \phi \pm B_m ~~.
\label{EM_eigenvectors}
\end{eqnarray}
The boundary conditions in this case are just given by
\begin{eqnarray}
  \partial_t (E_l - B_p) &=& 0 ~~,~~ \partial_t (E_p + B_l)=0 ~~,\\
  \partial_t (\Psi - E_m) &=& 0 ~~,~~ \partial_t (\phi - B_m)=0 ~~.
\label{EM_eigenvectors}
\end{eqnarray}

We adopt Finite Difference techniques on a regular Cartesian grid to implement
the overall system numerically. To ensure sufficient resolution is used in an efficient
manner we employ adaptive mesh refinement techniques. To this end we adopt the \had\ computational infrastructure 
that provides distributed Berger-Oliger
style Adaptive Mesh Refinement (AMR)~\cite{had_webpage,Liebling} with full sub-cycling
in time, together with a novel treatment of artificial boundaries~\cite{Lehner:2005vc}.
The refinement regions are determined using truncation error estimation via a shadow
hierarchy~\cite{Pretorius} and so they adapt dynamically as the evolution proceeds to guarantee a 
certain pre-specified tolerance is achieved. 
A fourth order spatial discretization satisfying a summation by parts rule
together with a third order Runge-Kutta scheme for the time integration are used to help
ensure stability of the numerical implementation~\cite{binaryNS}. We adopt a Courant
parameter of $\lambda = 0.2$ so that $\Delta t_l = 0.2 \Delta x_l$ at each level$=l$; this ensures
the implementation satisfies the Courant-Friedrichs-Levy condition dictated by the principal part of
the equations. However notice that the different damping terms (either at the gauge
condition (\ref{eq:harmonic_frans}) or constraint damping) in turn add a further
requirement of the form $\sigma\, \Delta t_l \simeq O(1)$. 
Since $\Delta t_l$ is considerably larger for the coarser grids
(where the solution is obtained at large distances) we address this issue by making
the different damping factors space dependent, and in particular 
\begin{equation}
\sigma_i = \left\{
\begin{array}{ll}
\hat \sigma_i & r \leq r_0 M \\
\hat \sigma_i e^{-(r-r_o M)^2/(10 M)^2}  & r > r_0 M 
\end{array}
\right.
\end{equation}
($i=0,1,2$).
For similar reasons we also consider a spatial dependence for $\xi_i$. While different options
work well, for comparison purposes we adopt a similar strategy as in \cite{Scheel:2008rj}, thus
\begin{equation}
\xi_i \rightarrow \hat \xi_i e^{-r^2/(40 M)^2} f_i \, ,
\end{equation}
with
\begin{equation}
f_1 = \left(2 - e^{\frac{t-t^*}{18M}} \right) \left(1 - e^{\frac{(t-t^*)^2}{(15M)^2}}\right) ~~ \, , ~~ \, f_2 = 1 \, ;
\end{equation}
with $t^*$ chosen to be $10M$ before the onset of merger. As mentioned, other options that
``turn on'' the gauge smoothly from a pure Harmonic condition to a generalized
one as given by eqn. (\ref{eq:harmonic_frans}) work as well.
 
%
%
\section{Physical setup and Initial data} 

We consider both single and binary black hole simulations,
immersed in an otherwise constant magnetic field like the one produced
by a disk surrounding the black hole at large distances. 
As the electromagnetic fields interact with the curved spacetime, they will be dynamically
distorted and eventually reach a quasi-stationary
configuration. We adopt electromagnetic fields within astrophysically
relevant values where their energy is several orders of
magnitude smaller than the gravitational field energy and so they have a negligible
influence on the black holes' dynamics. A simple estimate indicates this is the
case for a large (and certainly astrophysical interesting) range of field strengths.
To this end first we express the field in terms of units of $M^{-1}$
\begin{equation}
  B[1/M] = 1.2 \cdot 10^{-20} \left( \frac{M}{M_{\odot}} \right) B[G]
\end{equation}
so, fields up to $B_o=10^{18} (M_{\odot}/M) G$ will have energy densities $\leq 10^{-4} [M^{-2}]$.
Additionally, within a sphere of radius $\simeq 100M$, the total  EM energy will
be bounded by $\%1 M$ if the field strength is $\leq 10^{16} (M_{\odot}/M) G$.
Therefore, while we here adopt a field strength of $B_o = 10^{4} (M/10^8 M_{\odot})$~G, 
our results are applicable to much stronger values since for fields up 
to $\simeq 10^{16} (M_{\odot}/M) G$ the 
effects of the electromagnetic fields on the geometry are negligible however
the latter can have a profound effect on the former. 

The analysis of the single black hole case will serve not only as a test for our
numerical implementation, but also to understand the features of the initial
transient, where the EM fields adapt to the geometry of the black hole spacetime,
giving rise to an electric field and a deformation of the magnetic field 
(see also~\cite{KiLaKu75,Komissarov:2007rc}).
As it was shown in \cite{KiLaKu75}, the quasi-stationary state is determined by 
Wald's solution~\cite{1974PhRvD..10.1680W} for a Kerr black hole immersed in a uniform
magnetic field which is aligned with its spin. Let us consider the explicit form of
Wald's solution, which describes a solution of the 
Maxwell equations in the test field case. We assume a Kerr black hole in Boyer-Lindquist (BL) coordinates, immersed in a uniform magnetic field \cite{1974PhRvD..10.1680W}.

\begin{eqnarray}
  ds^2 &=& - \left(1 - \frac{2 M r}{\Sigma}\right) dt^2 
           - \frac{4 M a r \sin^2 \theta}{\Sigma} dt d\phi 
\\
       &+& \left[\frac{(r^2+a^2)^2 - \Delta a^2 \sin^2\theta}{\Sigma} \right] \sin^2\theta d\phi^2
       + \frac{\Sigma}{\Delta} dr^2 + \Sigma d\theta^2
\nonumber
\end{eqnarray}
where $\Sigma = r^2 + a^2 \cos^2\theta$ and $\Delta = r^2 + a^2 - 2 M r$.
On this background, the following tensor defines a solution of Maxwell equations:
\begin{eqnarray}
    F &=& F_{10}\, \omega^1 \wedge \omega^0 + F_{13}\, \omega^1 \wedge \omega^3
    + F_{20}\, \omega^2 \wedge \omega^0 
\nonumber \\
    &+& F_{23}\, \omega^2 \wedge \omega^3  \, ,
\end{eqnarray}
where $\omega^a \wedge \omega^b \equiv \frac{1}{2} (\omega^a \otimes \omega^b
 - \omega^b \otimes \omega^a)$, with
\begin{eqnarray}
    \omega^0 &=& \left(\frac{\Delta}{\Sigma} \right)^2 \left(dt - a \sin^2\theta d\phi \right) ~~,~~
    \omega^1 = \left( \frac{\Sigma}{\Delta} \right)^2 dr ~~,~~
\nonumber \\
    \omega^2 &=& \Sigma^{1/2} d\theta ~~,~~
    \omega^3 = \frac{\sin\theta}{\Sigma^{1/2}}\left[(r^2+a^2)d\phi - a dt \right] 
\\
\nonumber \\
    F_{10} &=& B_0 \left[ \frac{a r \sin^2\theta}{\Sigma} 
                - \frac{M a (r^2 - a^2 \cos^2\theta)(1 + \cos^2\theta)}{\Sigma^2} \right] ~~,~~
\nonumber \\
    F_{13} &=& B_0 \frac{\Delta^{1/2} r \sin\theta}{\Sigma} ~~,~~
    F_{20} = B_0 \frac{\Delta^{1/2} a \sin\theta \cos\theta}{\Sigma} ~~,~~
\nonumber \\
    F_{23} &=& B_0 \frac{\cos\theta}{\Sigma} \left[r^2 + a^2 - \frac{2 M r a^2 (1 + \cos^2\theta)}{\Sigma} \right] \, .
\end{eqnarray}
Here $B_0$ is the magnitude of the magnetic field at large distances from the black hole.
For convenience we write the solution's components in the standard coordinates $\{t,r,\theta,\phi\}$,
\begin{eqnarray}
  F &=& F_{rt}\, \omega^r \wedge \omega^t + F_{r \phi}\, \omega^r \wedge \omega^\phi 
    + F_{\theta t}\, \omega^{\theta} \wedge \omega^t 
\nonumber \\
   &+& F_{\theta \phi}\, \omega^{\theta} \wedge \omega^{\phi} \, ,
\\
\nonumber \\
  F_{rt} &=& F_{10} - \frac{a \sin\theta}{\Delta^{1/2}}\, F_{13} ~~,~~
  F_{\theta t} = \Delta^{1/2}\, F_{20} - a \sin\theta\, F_{23} ~~,~~
\nonumber \\
  F_{r \phi} &=& - a \sin^2\theta\, F_{10} + \frac{\left(r^2 + a^2\right)}{\Delta^{1/2}}
     \sin\theta \, F_{13} \, ,
\nonumber \\
  F_{\theta \phi} &=& -a \Delta^{1/2} \sin^2\theta\, F_{20} 
        + \left(r^2+a^2\right) \sin\theta \, F_{23} ~~.
\end{eqnarray}

Lastly, since we employ excision techniques we prefer to adopt horizon penetrating coordinates and so
transform this solution to Kerr-Schild coordinates by the transformation~\cite{Komissarov:2004b}
\begin{eqnarray}
  &dt& \rightarrow dt - \frac{2 r}{\Delta} dr \, ,
\nonumber \\
  &d\phi& \rightarrow d\phi - \frac{a}{\Delta} dr \, .
\end{eqnarray}
The explicit expressions of the fields in these coordinates are lengthy but straightforward.  
To gain some
 insight on the solution however, the particular case of 
a non-spinning black hole suffices (setting $a=0$).
The line element in the BL coordinates reduces to
\begin{eqnarray}
  ds^2 &=& - \left(1 - \frac{2 M}{r} \right) dt^2 
         + \left(\frac{r^2}{r^2 - 2 M r} \right) dr^2 
\nonumber \\
         &+& r^2 d\theta^2 + r^2 \sin^2\theta d\phi^2 ~~.
\end{eqnarray}
The electric field vanishes ($E^{i}=0$) and the magnetic field components are given by
\begin{equation}
   B^{r} = B_0\, \alpha\, \cos\theta~,~
   B^{\theta} = - B_0 \frac{\alpha\, \sin\theta}{r} ~,~
   B^{\phi} = 0 ~.
\end{equation}

The corresponding expressions in Kerr-Schild coordinates are:
\begin{eqnarray}
   B^{r} &=& B_0\, \alpha\, \cos\theta~,~
   B^{\theta} = - B_0 \frac{\alpha\, \sin\theta}{r}~,~
   B^{\phi} = 0~, \\
   E^{r} &=& E^{\theta} = 0 ~,~
   E^{\phi} = \frac{2\, B_0\, \alpha\, M}{r^2} ~.
\end{eqnarray}

Notice that although the magnetic field has the 
same expression in both systems of coordinates, there appears a toroidal component in the 
electric field in these coordinates due to the non-vanishing shift vector.\\

To explore the effects of the merger dynamics on the electromagnetic field,
we compare single spinning black hole with cases of equal-mass merging black holes.
In all cases, the orbital plane of the evolution (or equatorial plane for the single BH)
is assumed to be aligned with that of the circumbinary disk. The magnetic field
is defined as anchored in the disk; hence, its associated magnetic dipole is aligned
with the orbital and spin angular momentum.

%
%
\section{Single black holes; the asymptotic stationary state}

For the single black hole case, we adopt a spinning black hole with a spin
parameter given by $a=0.7M$ which is close to the spin expected for a merged black hole 
from an  equal-mass, non-spinning binary system. We adopt this value for comparison
with the binary black hole scenario presented in the next section, where 
the final spin can be calculated directly via simulations 
(for recent efforts in simulations and data analysis 
see e.g.~\cite{Aylott:2009ya} and references cited therein) 
or estimated by simple arguments as in~\cite{Buonanno:2007sv}.
In this simulation the geometry is kept fixed, in order to maintain the
same initial choice of coordinates and be able to compare easily with 
 Wald's (analytical) solution. 

As mentioned, the initial magnetic field is described by a poloidal configuration constructed 
from the electromagnetic potential produced by a circular loop, whose radius is assumed
to be larger than the region of interest~\cite{Jackson1975}. We assume the disk lies
at $10^3M$, and for these distances the magnetic field is essentially constant within our
computational domain, so 
we simply adopt $B^i = B_o \hat z$ and set 
the electric field initially to zero throughout.
The magnetic field strength is $B_o = 10^{4}$~G,
which is consistent with possible values inferred in relevant astrophysical systems
\cite{2008A&A...477....1M,1993ApJ...403...94F}. 

Our numerical domain consists of a cubical region defined by
$[-80 \,M, 80 \,M]^3$ with $80$ points in the base grid. It
employs an FMR configuration with 
$5$ levels of refinement, each one covering half of the domain of the parent
coarser level. Thus, the coarsest resolution employed is $\Delta x=2M$
while the finest one is $\Delta x=0.125M$. The damping parameter is
set to be $\hat \sigma_2 = 1M$.

\begin{figure}[h]
\begin{center}
\epsfig{file=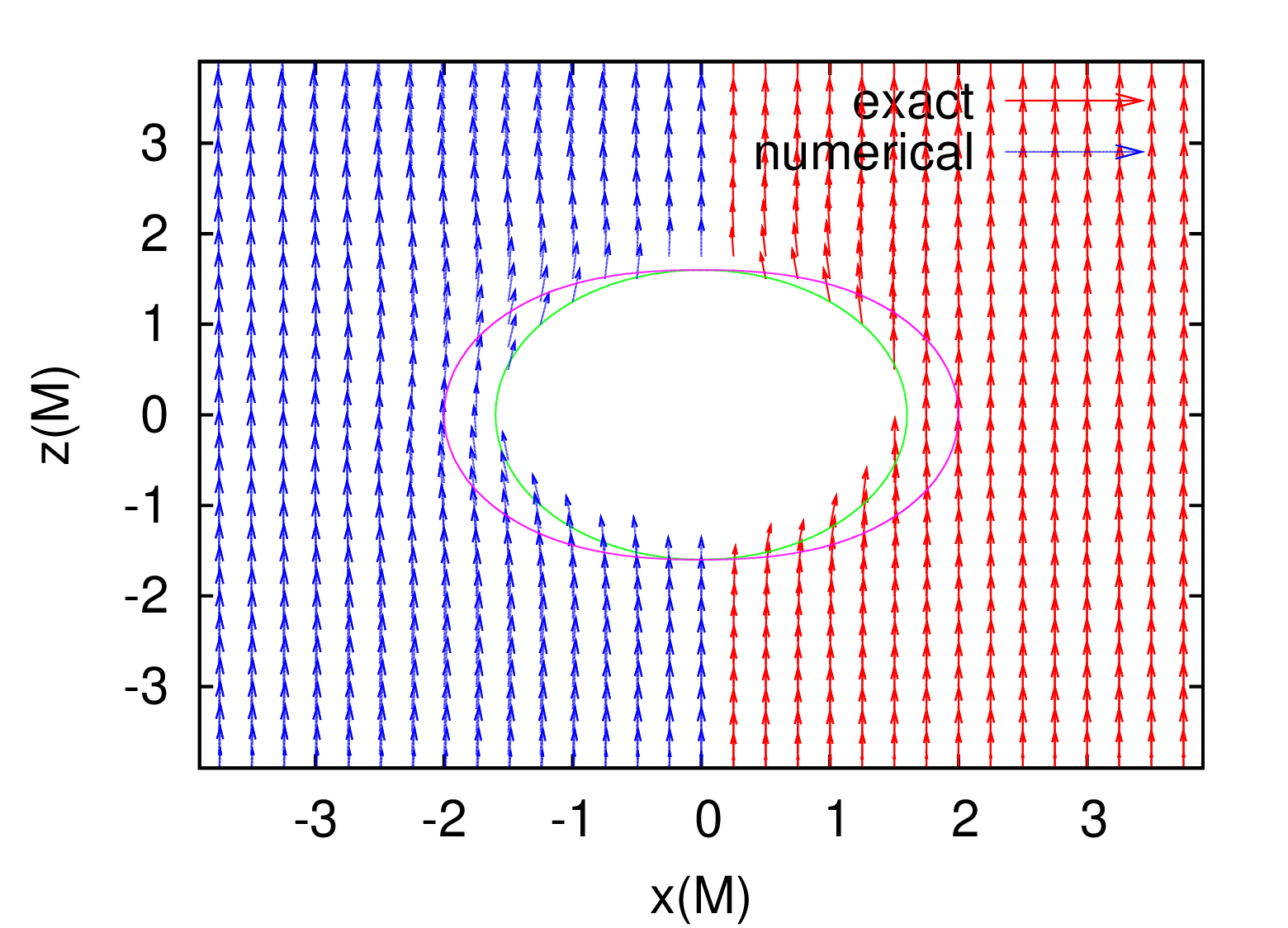,height=4.6cm,width=6.2cm}
\epsfig{file=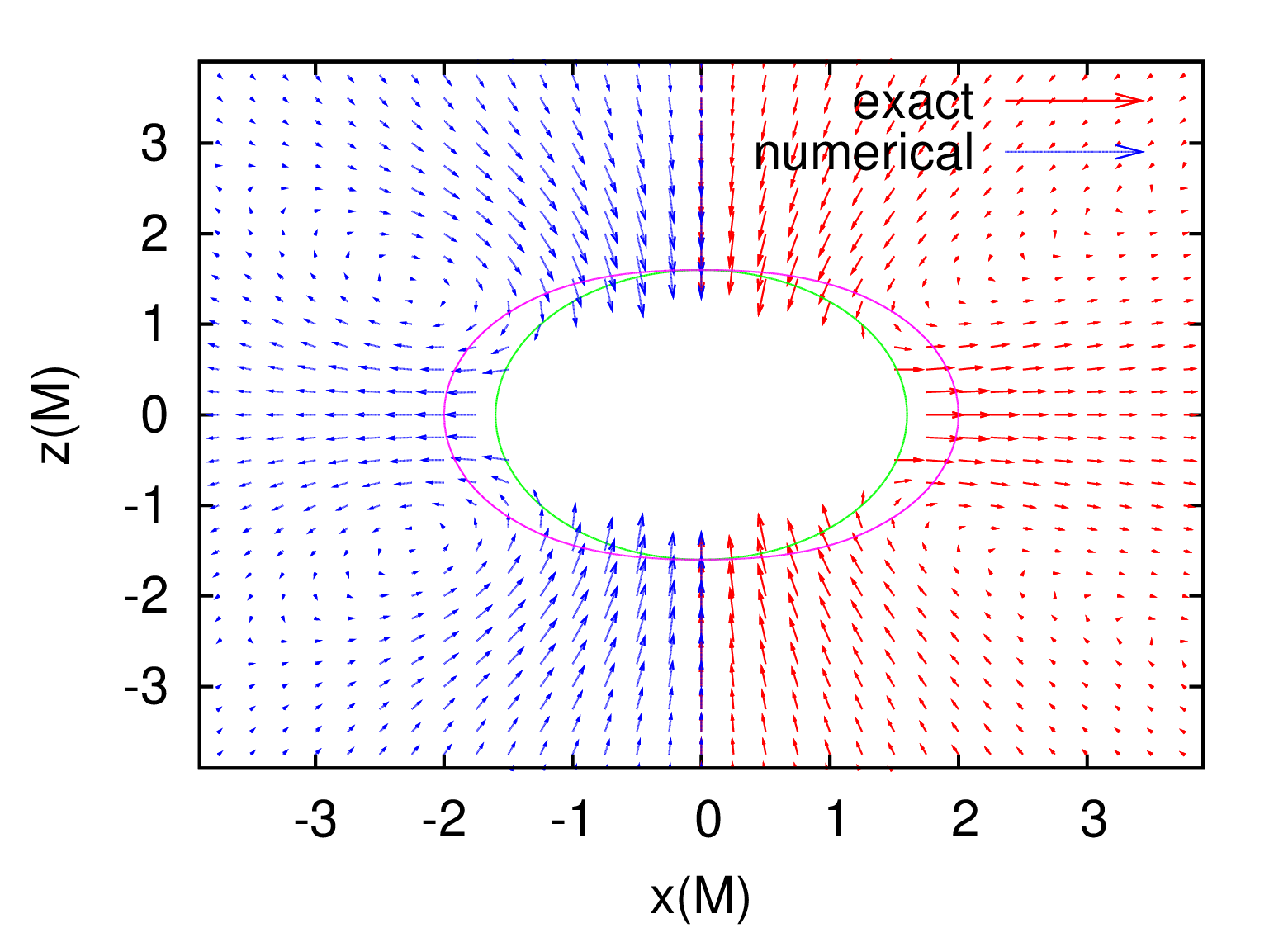,height=4.6cm,width=6.2cm}
\caption{Magnetic and electric field lines for $y=0,z\geq 0$ at $t=200M$ for a
single black hole, together with the apparent horizon~(green) and the ergosphere~(magenta).
The regions $x<0$ display the numerical solution, while Wald's exact solution is shown in $x>0$.}
\label{fig:single_field}
\end{center}
\end{figure}

The evolution shows an initial transient where the magnetic fields are deformed through
the dynamics exhibit a twisting behavior around the spinning black hole as well as an
induced electric field.
After $t \simeq 80M$ the solution is clearly seen to evolve towards a quasi-stationary state 
determined by Wald's solution~\cite{1974PhRvD..10.1680W} for a Kerr black hole immersed in a uniform
magnetic field which is aligned with its spin. This is illustrated in Fig.~\ref{fig:single_field} which presents
both the electric and magnetic field obtained at $t=200M$ in
the plane $y=0$ for $x>0$, the corresponding field from Wald's exact solution is shown for $x>0$. 
The apparent agreement along with a careful examination of the asymptotic
solution indicates that, for a black hole immersed within an almost uniform magnetic field aligned with its spin, 
the final state is  Wald's solution~\cite{KiLaKu75}.

It is interesting to notice that the electric field configuration is equivalent
to the one produced by a sphere with a surface density charge, inmersed in a external
magnetic field. This is precisely the result expected within the membrane paradigm picture, 
which endows the event horizon with some physical properties (ie, temperature, current, density charge,etc.) \cite{1986bhmp.book.....T}. From this point of view, the density charge distribution is thus responsible for the electromagnetic field configuration observed. 

%
%
\section{Binary black holes and effects on EM fields}
We now turn our attention to the binary black hole case. As argued earlier the
electromagnetic field considered (which is below estimated upper bounds in astrophysical scenarios)
is not strong enough to affect the dynamics of the black holes, rather the black holes
will affect the electromagnetic fields. Prior to merger, when the black holes
are far from each other, the physical picture of the EM fields' behavior can be intuitively
obtained from the knowledge of the black holes quasi-circular trajectories and the 
membrane paradigm point of view together with results from the previous section.
Namely the black holes immersed in a uniform field affect the electromagnetic field's configuration 
in its local neighborhood, and the trajectories of the black holes cause a charge separation in the direction
perpendicular to both the velocity and magnetic field, as in the Hall effect. As a result, the electromagnetic
field induced by the binary motion is given by two dipoles in a quasi-adiabatical shrinking orbital behavior.
As the merger stage approaches, the strong curvature and dynamics might affect the EM field's behavior
more strongly and so we concentrate on this stage. We adopt initial data such that the merger takes place after about one orbit.
This initial data corresponds to quasi-equilibrium, equal-mass, non-spinning black holes
constructed by the publicly available {\sc lorene} code~\cite{lorene_webpage}.
The black holes have masses,
given by $M_s = M/2$, and are initially separated by $\approx 6 M$, lying beyond
the approximate inner most stable circular orbit (ISCO)~\cite{Buonanno:2007sv}. 
The initial magnetic field is chosen, as in the single black hole case,
to be a poloidal configuration produced by a circular loop with large radius,
so that $B^i = B_o \hat z$. The electric field is initially zero throughout
the computational domain and the magnetic field strength adopted  
is $B_o = 10^{4}$~G.

\begin{figure}[h]
\begin{center}
\epsfig{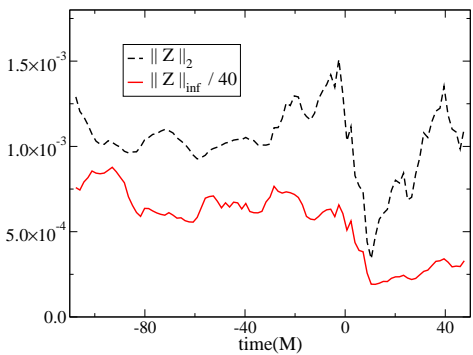}
\caption{$L_2$ and the $L_{\infty}$ norms of the physical constraint $||Z||$, which is
kept under control through the evolution with a small increase at the merger. After it,
the norms grow and oscillate for some period as the finest grid is detroyed and created
a few times as required by the tolerance adopted at times $> 38M$ the finest is destroyed
for the last time and not needed anymore.
}
\label{fig:physZ}
\end{center}
\end{figure}

We adopt a cubical domain given by $[-106 \,M, 106 \,M]^3$ and employ an AMR configuration with 
$6$ levels of refinement that adjust themselves dynamically to ensure that
the solution's error is below a pre-determined threshold using
a shadow hierarchy. The coarsest grid has $46$ points, so the coarsest
resolution is $\Delta x =4.6 M$ near the boundaries while the finest one is
$\Delta x=0.072 M$ around the black holes. We adopt the following set
of gauge parameters $\hat \xi_1 =0.084/M^2$, $\hat \xi_2 = 9/M$ and $n=3$, while the
damping parameters are $\hat \sigma_0=\hat \sigma_1= \hat \sigma_2=0.25/M$.
We monitor that the constraint remain well behaved
through the evolutions; for instance fig.~\ref{fig:physZ} displays the 
$L_2$ and the $L_{\infty}$ norms of the physical constraint $||Z||$ (\ref{constraints}), defined as
\begin{equation}
  ||Z|| \equiv \displaystyle\sum_{a=0}^{3} Z_a^2 ~~.
\end{equation}
Notice that besides a small increase at the time of the merger and at late times
when the finest grid is automatically discarded and recreated a few times automatically
as required by the shadow hierarchy, this constraint (as well as all others) remains under control
during simulation.

\begin{figure}[h]
\begin{center}
\epsfig{file=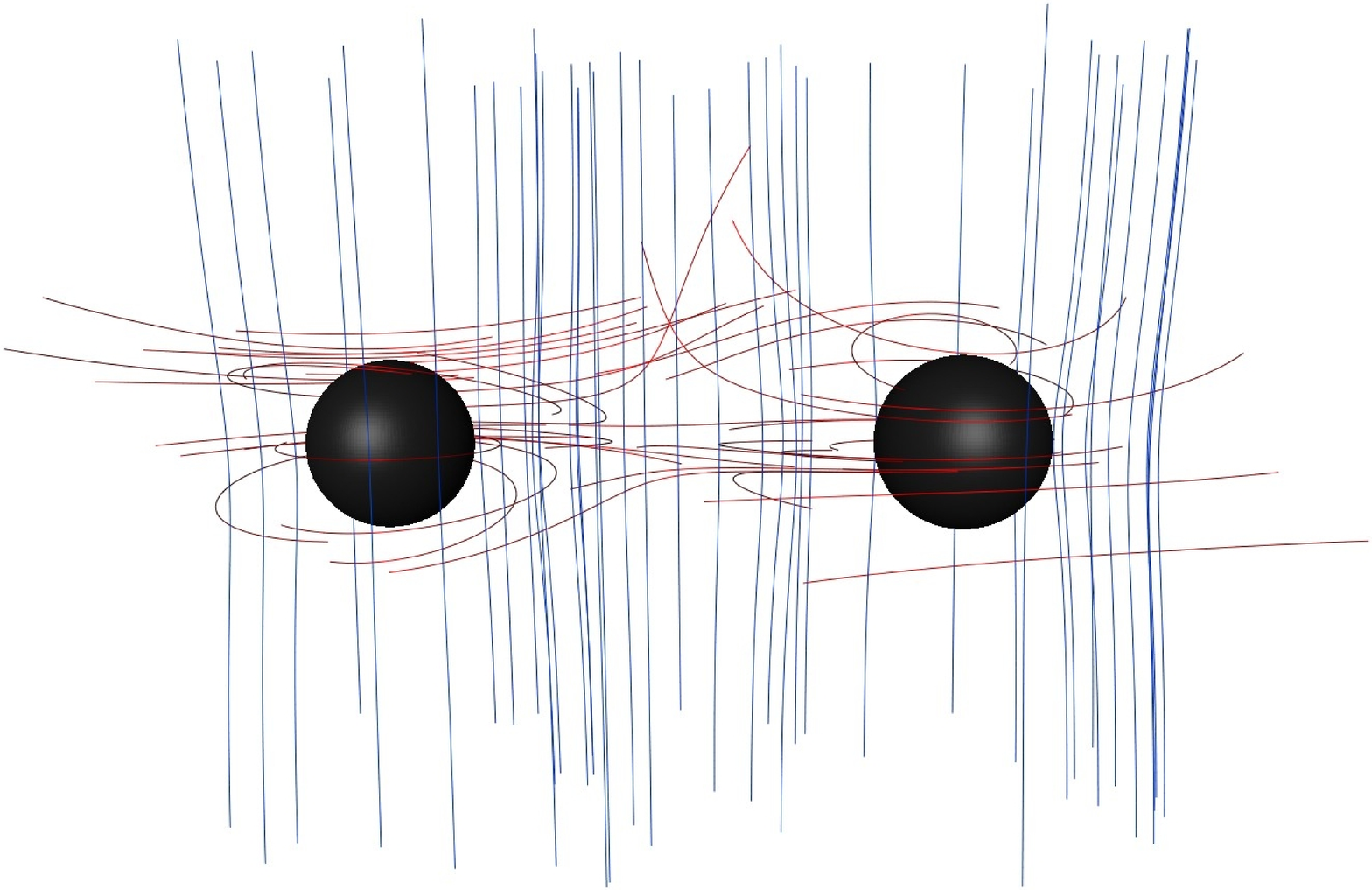,height=4.2cm,width=4.2cm}
\epsfig{file=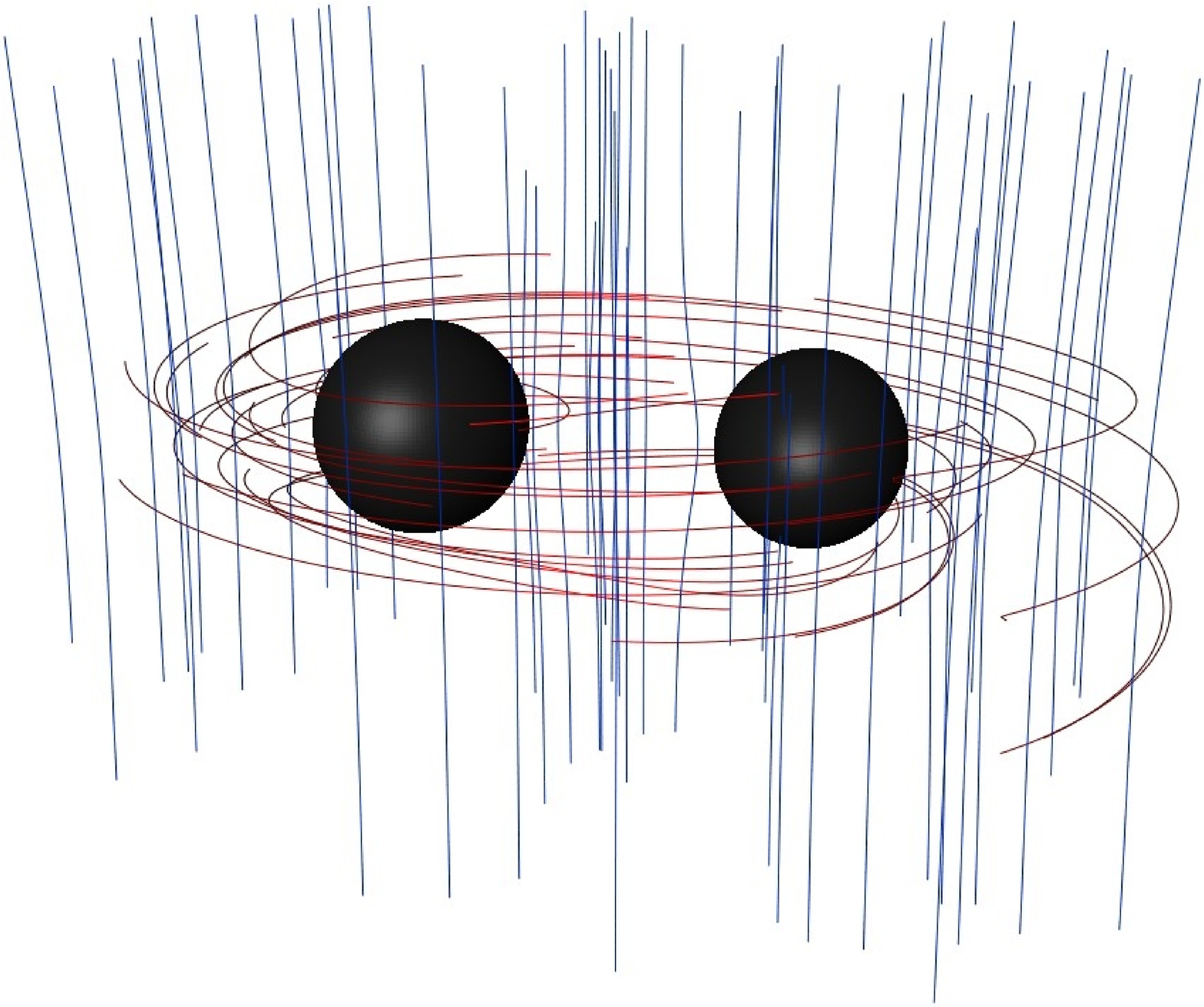,height=4.2cm,width=4.2cm}
\epsfig{file=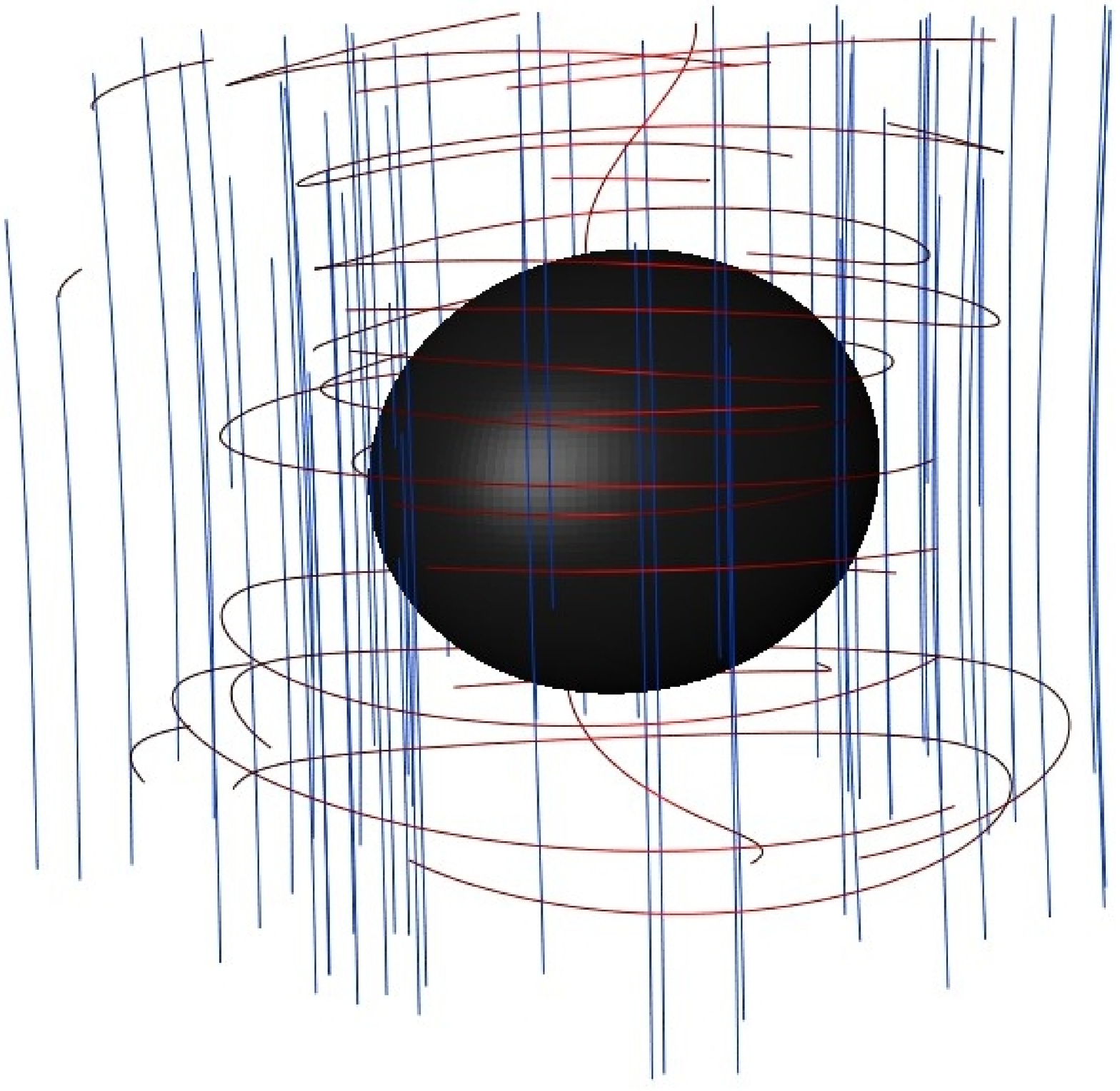,height=4.2cm,width=4.2cm}
\epsfig{file=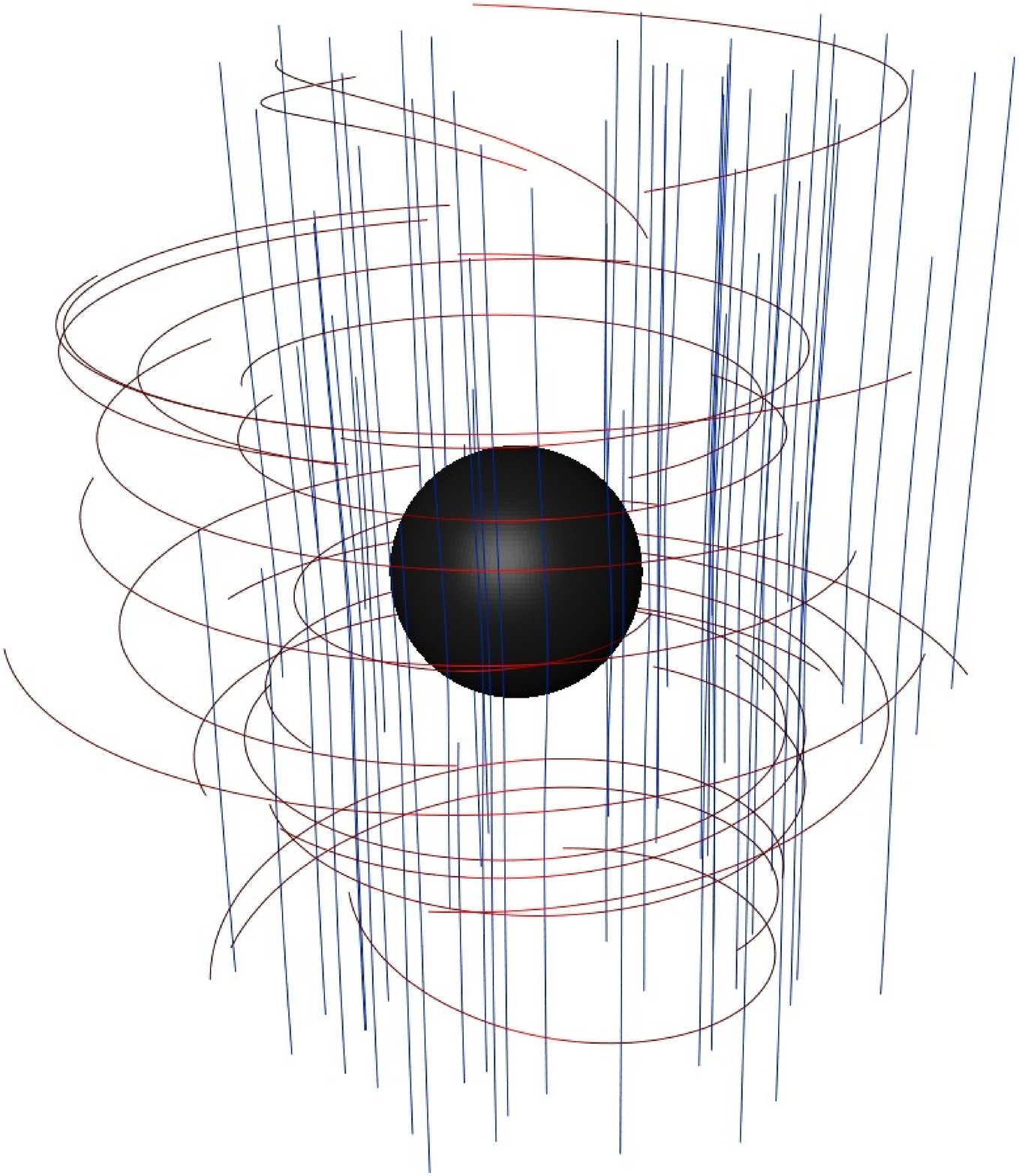,height=4.2cm,width=4.2cm}
\caption{Magnetic (mostly vertical) and electric field lines at different phases 
during the evolution employing different scales for visualization purposes. The
figures illustrate different stages: early when the
black holes are separated; near merger; shortly after they merge and
at late times. The electric field lines are twisted around the black hole, while the 
magnetic lines slightly deform from ther initial configuration aligned with the $z$-axis.}
\label{fig:binary_3dfield}
\end{center}
\end{figure}

A careful inspection of the dynamics described by the numerical simulation
revealas that, except at late times after merger, the physical behavior results significantly different
than the one corresponding to the single black hole case. Indeed the orbiting black holes modify both 
the geometry and the electromagnetic fields. As a result, the EM fields are stirred during 
the evolution, changing their configurations as it is displayed in Fig.~\ref{fig:binary_3dfield}.

\begin{figure}[h]
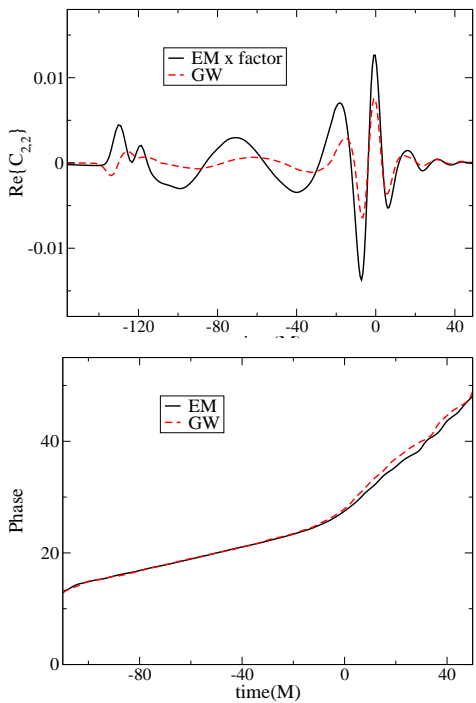

\begin{center}
\epsfig{file=NPscalars2.eps,height=4.6cm,width=6.2cm}
\epsfig{file=phase22.eps,height=4.6cm,width=6.2cm}
\caption{The top figure corresponds to the $l=m=2$ modes for $r \Psi_4$ and 
$r \Phi_2$ extracted at $r=40M$, reescaled properly with a factor $\approx 3 \times
10^6$ to fit in the same scale. The bottom figure illustrates the phase of these modes.}
\label{fig:NPscalars2}
\end{center}
\end{figure}

To analyze the influence of the binary's dynamics on the electromagnetic field we monitor
the (EM) Newman-Penrose radiative quantity $\Phi_2$, and correlate it with $\Psi_4$,
which is the gravitational wave Newman-Penrose scalar. These scalars are computed
by contracting the Maxwell and the Weyl tensor respectively, with a suitably defined
null tetrad 
\begin{eqnarray}
  \Phi_2 = F_{ab} n^a \bar m^b ~~,~~ \Psi_4 = C_{abcd} n^a \bar m^b n^c \bar m^d ~~;
\end{eqnarray}
extracted at a sphere surface $\Sigma$ located in the wave-zone,
far away from the sources. We also check that corrections required for possible
gauge issues, as discussed in \cite{Lehner:2007ip}, are negligible in our present case. 
To understand the induced multipolar structure of these quantities we
decompose them in terms of spin-weighted spherical harmonics, with spin weight $s=-1$
for $\Phi_2$ and $s=-2$ for $\Psi_4$ (since these are their respective spin-weights).
These modes exhibit a very similar behavior, with the most relevant ones corresponding to
the $l=2,m=\pm 2$ modes, which are plotted in fig.~\ref{fig:NPscalars2} (top). The maximum amplitudes
of these waveforms correspond to the merger time which takes places after over one orbit.
Note that since the magnetic field is anchored at the disk it does not decay with 
the distance from the binary, which obscures a clean interpretation from $\Phi_2$, displaying 
a non-vanishing $m=0$ mode at late times, when the stationary state is reached. The same happens
 with the decomposition of the radial component of the Poynting vector $S_r$ (radial from the origin),
 which shows non-radiative modes not related to the binary black hole dynamics.
A closer inspection of the waveforms (bottom in fig.~\ref{fig:NPscalars2}) reveals that the $l=m=2$
mode of both the GW and the EM waves oscillate with the same frequency, indicating that both
are mostly dominated by a quadrupolar structure resulting from the orbiting behavior.

\begin{figure}[h]
\begin{center}
\epsfig{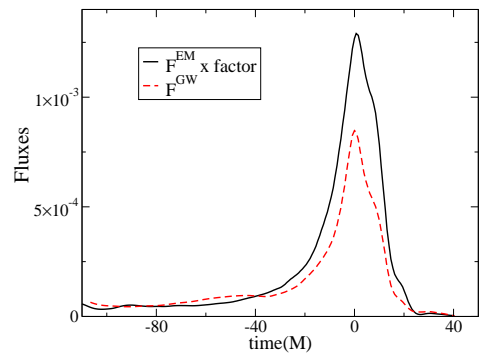}
\caption{Total flux of gravitational and electromagnetic energy corresponding to the
binary black hole case 
(integrated over the sphere surface $\Sigma$). The (integrade) electromagnetic energy flux
has been reescaled by a factor $\approx 10^{13}$ to make it appear clearly in the plot.}
\label{fig:flux}
\end{center}
\end{figure}

The energy carried off by outgoing waves at infinity is another interesting quantity.
The total energy flux per unit solid angle can be found directly from the
Newman-Penrose scalars. 
\begin{eqnarray}
  F^{GW} &=& \frac{{dE}^{GW}}{dt\, d\Sigma} = 
\lim_{r \rightarrow \infty} \frac{r^2}{16 \pi} \left| \int_{\infty}^t \Psi_4 dt' \right|^2 \, ,
\label{FGW} \\
  F^{EM} &=& \frac{{dE}^{EM}}{dt\, d\Sigma} = \lim_{r \rightarrow \infty} 
                                       \frac{r^2}{2 \pi} |\phi_2|^2 ~~.
\label{FEM}
\end{eqnarray}
These quantities, integrated along the sphere surface $\Sigma$,
located at $R_{\Sigma} = 30 M$  are shown in fig.~\ref{fig:flux}, both exhibiting a maximum 
at the time of the merger. 

\begin{figure}[h]
\begin{center}
\epsfig{file=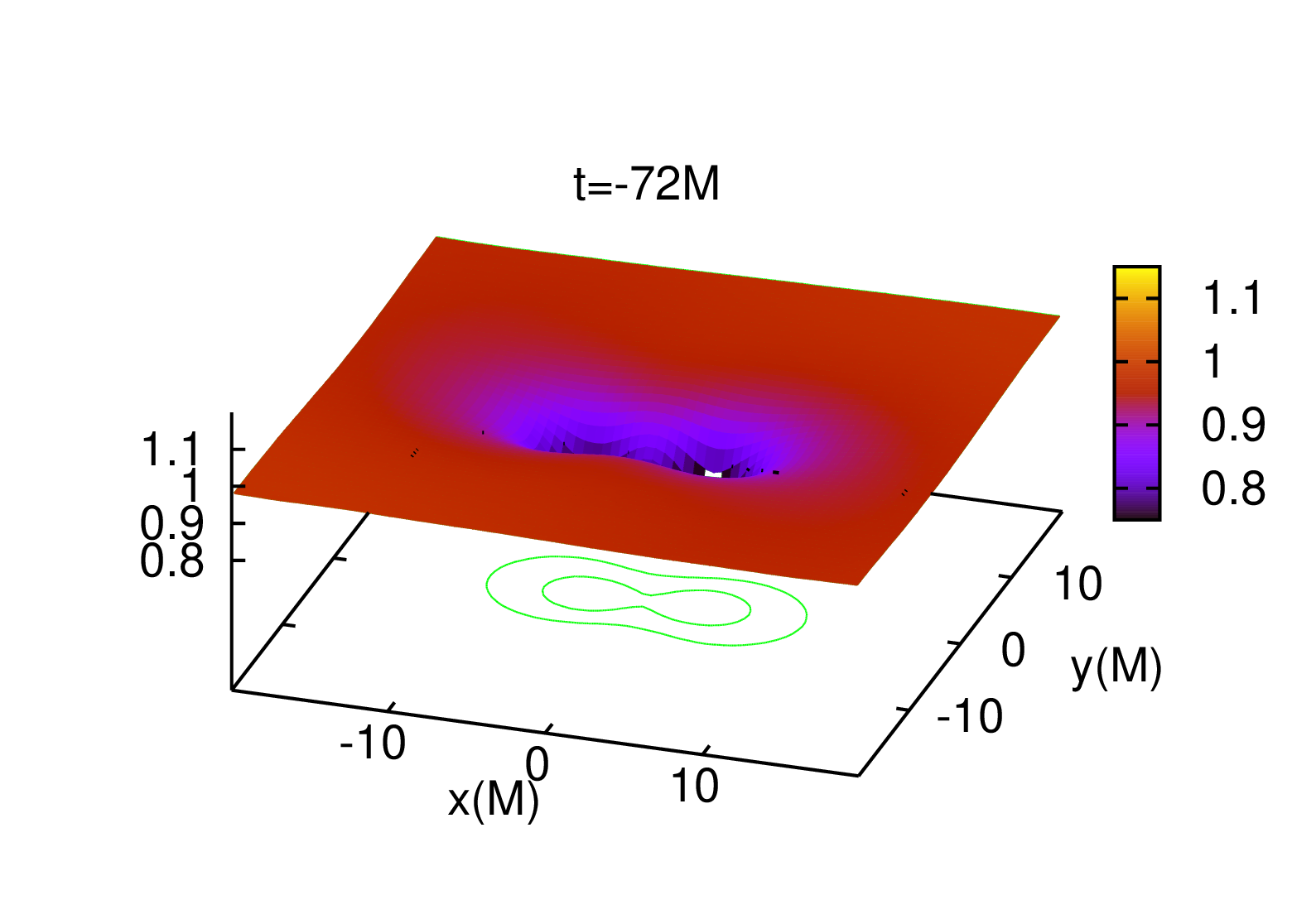,height=3.5cm,width=4.2cm} 
\epsfig{file=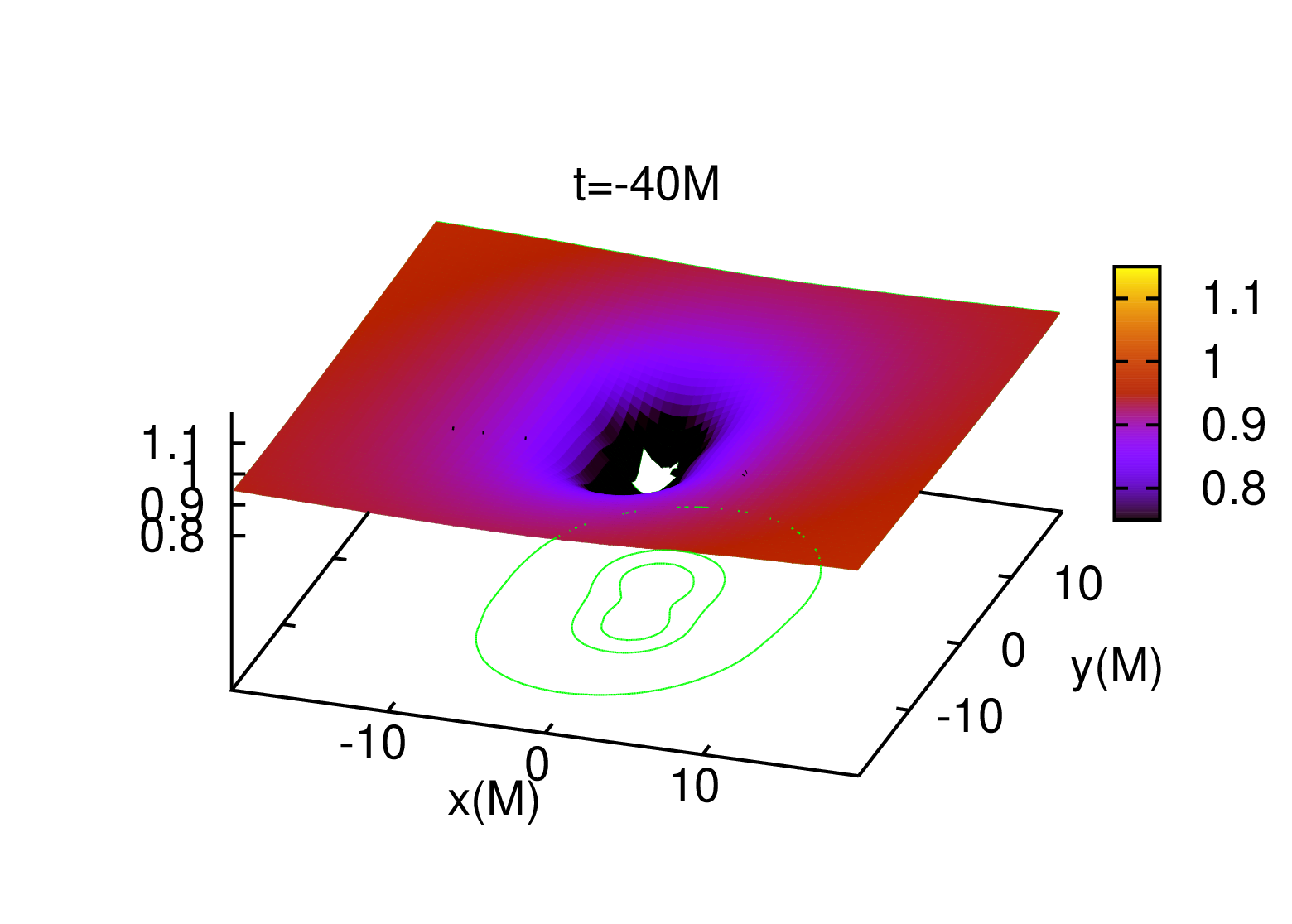,height=3.5cm,width=4.2cm} 
\epsfig{file=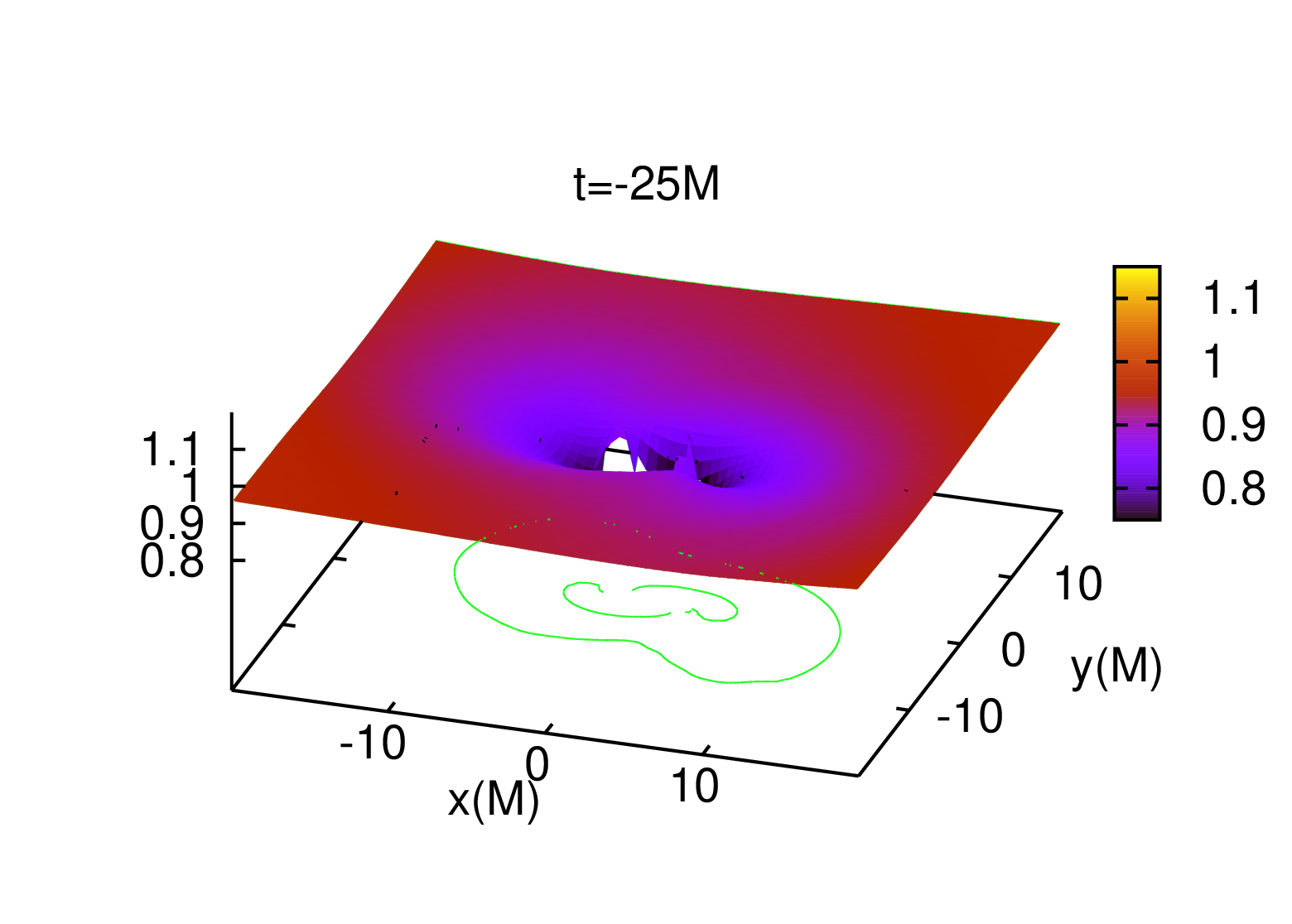,height=3.5cm,width=4.2cm} 
\epsfig{file=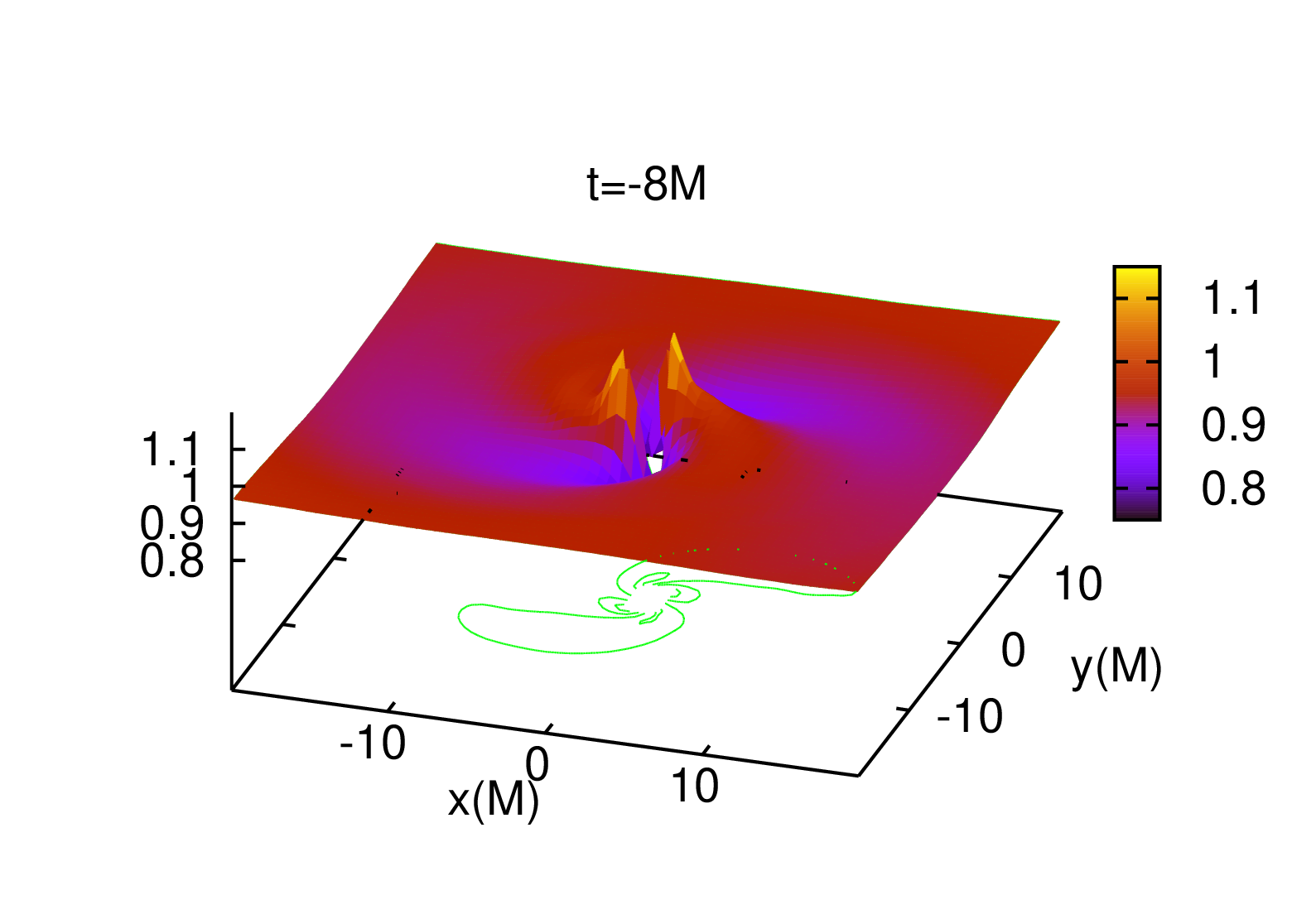,height=3.5cm,width=4.2cm} 
\epsfig{file=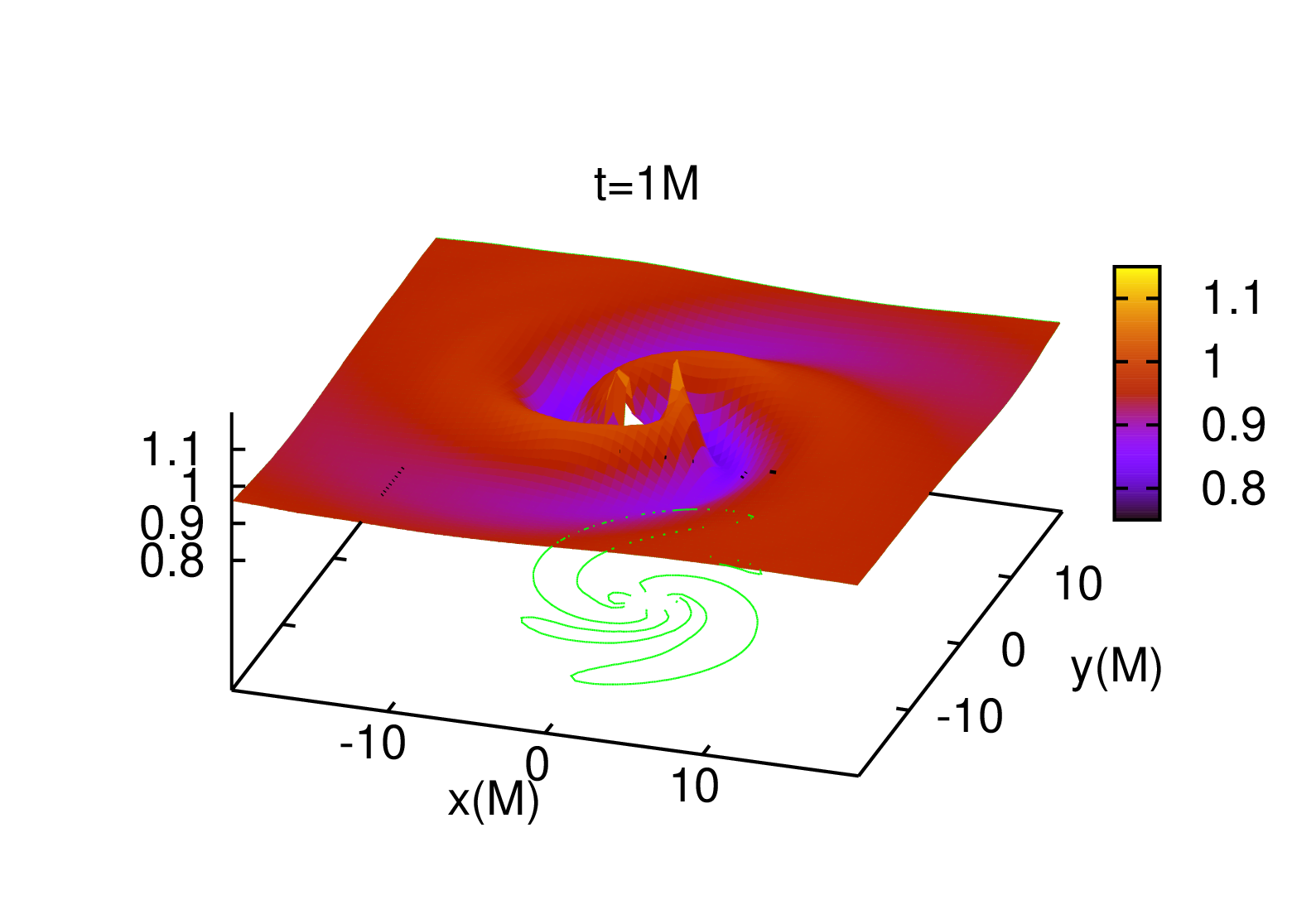,height=3.5cm,width=4.2cm} 
\epsfig{file=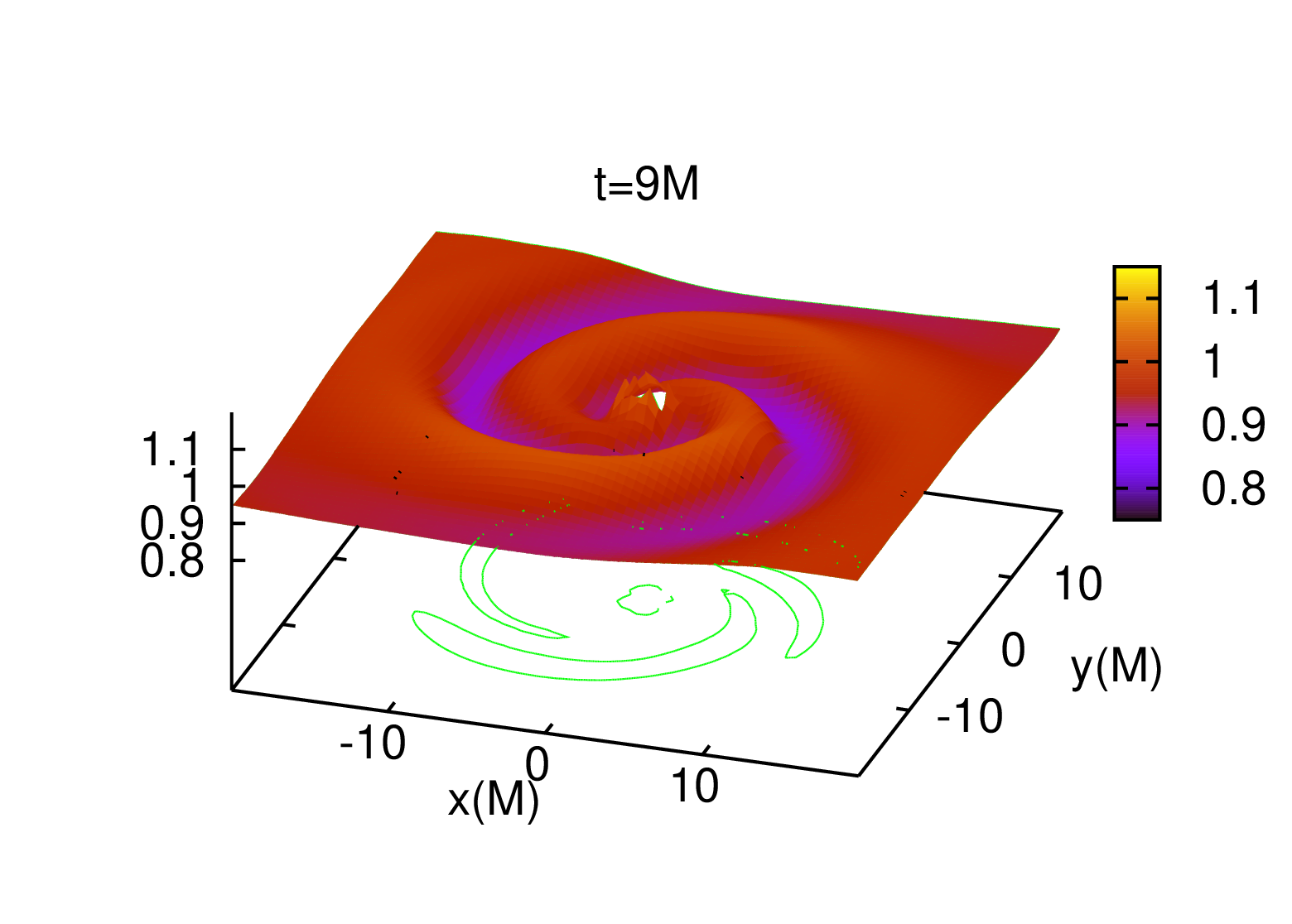,height=3.5cm,width=4.2cm} 
\epsfig{file=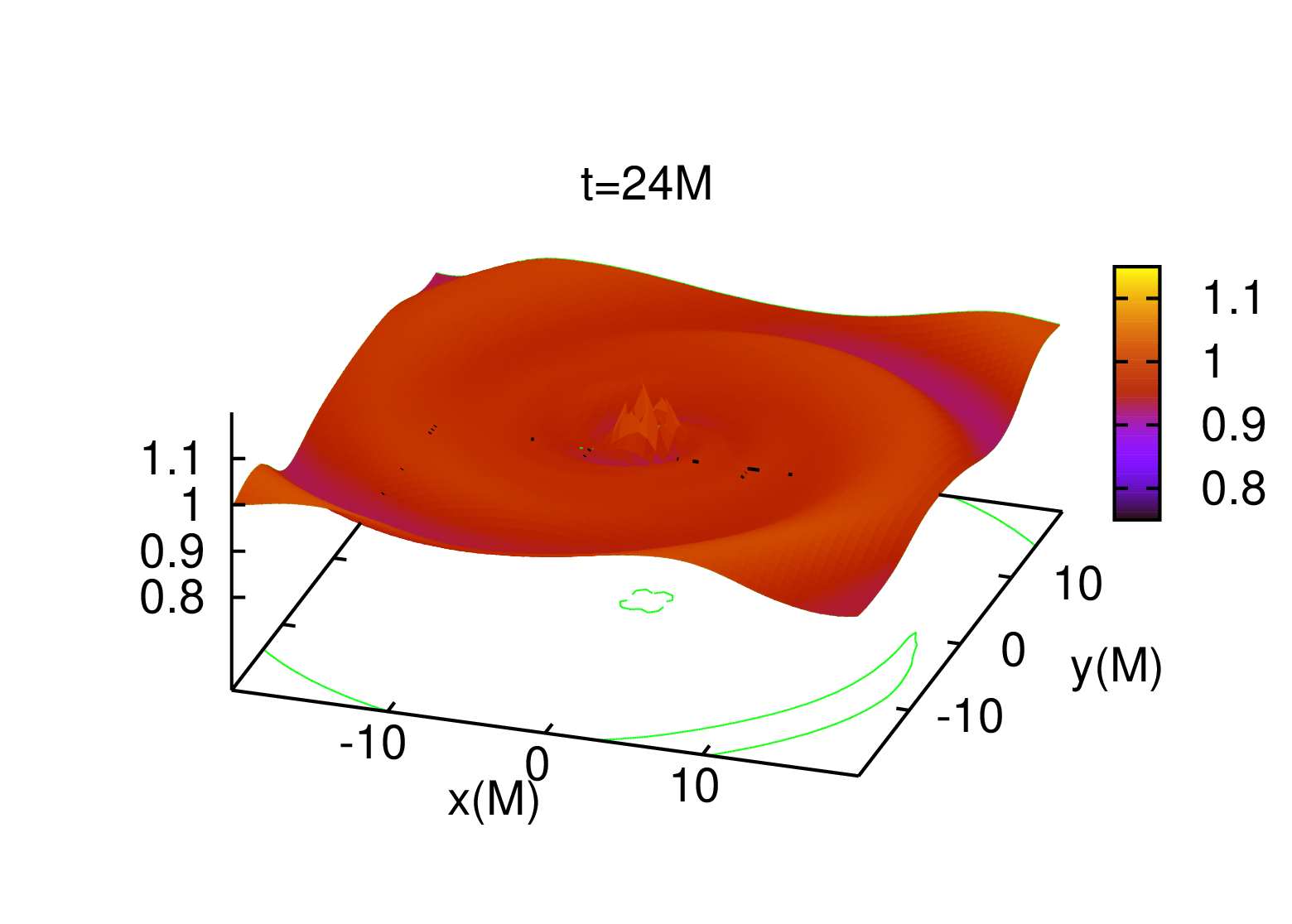,height=3.5cm,width=4.2cm} 
\epsfig{file=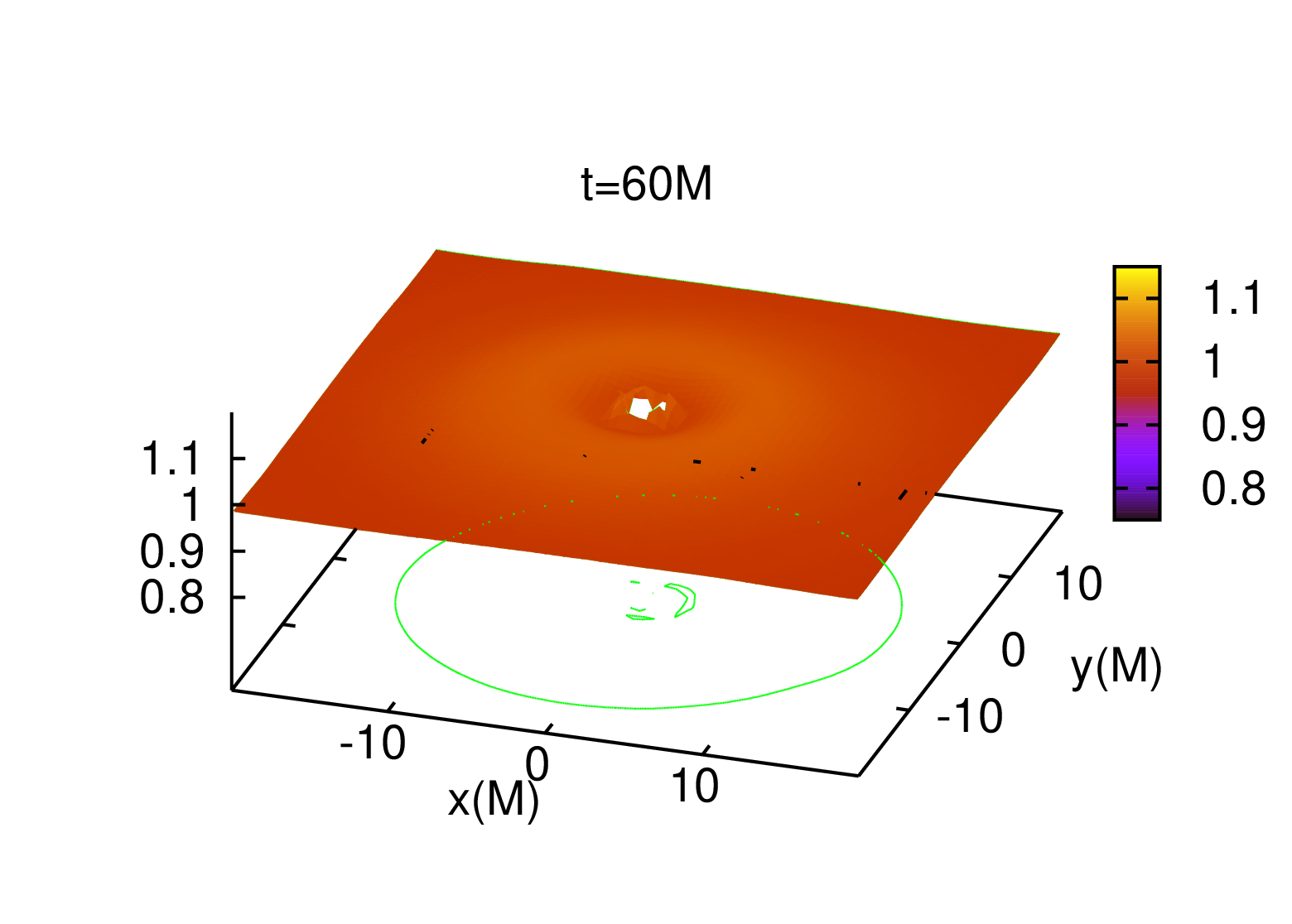,height=3.5cm,width=4.2cm} 
\caption{Electromagnetic energy density (normalized to the values at the initial time, 
far distances) 
at the $z=0$ plane for the binary black hole case
at different times of the evolution. Energy contour plots are shown from $1.2$ to $0.8$
at intervals of $0.1$.}
\label{fig:binary_B}
\end{center}
\end{figure}

Furthermore, the electromagnetic energy density ${\cal E}_{EM} \equiv (E^2 + B^2)/2$,
while not an invariant quantity in general relativity can be employed to get
a sense of the energy variation in the EM field.
As it is illustrated in Fig.~\ref{fig:binary_B}, as the merger takes
place, ${\cal E}_{EM}$ grows significantly during the merger and then diminishes as it
approaches its asymptotic stage which is described by the scenario described in the
single black hole case.

\begin{figure}[h]
\begin{center}
\epsfig{file=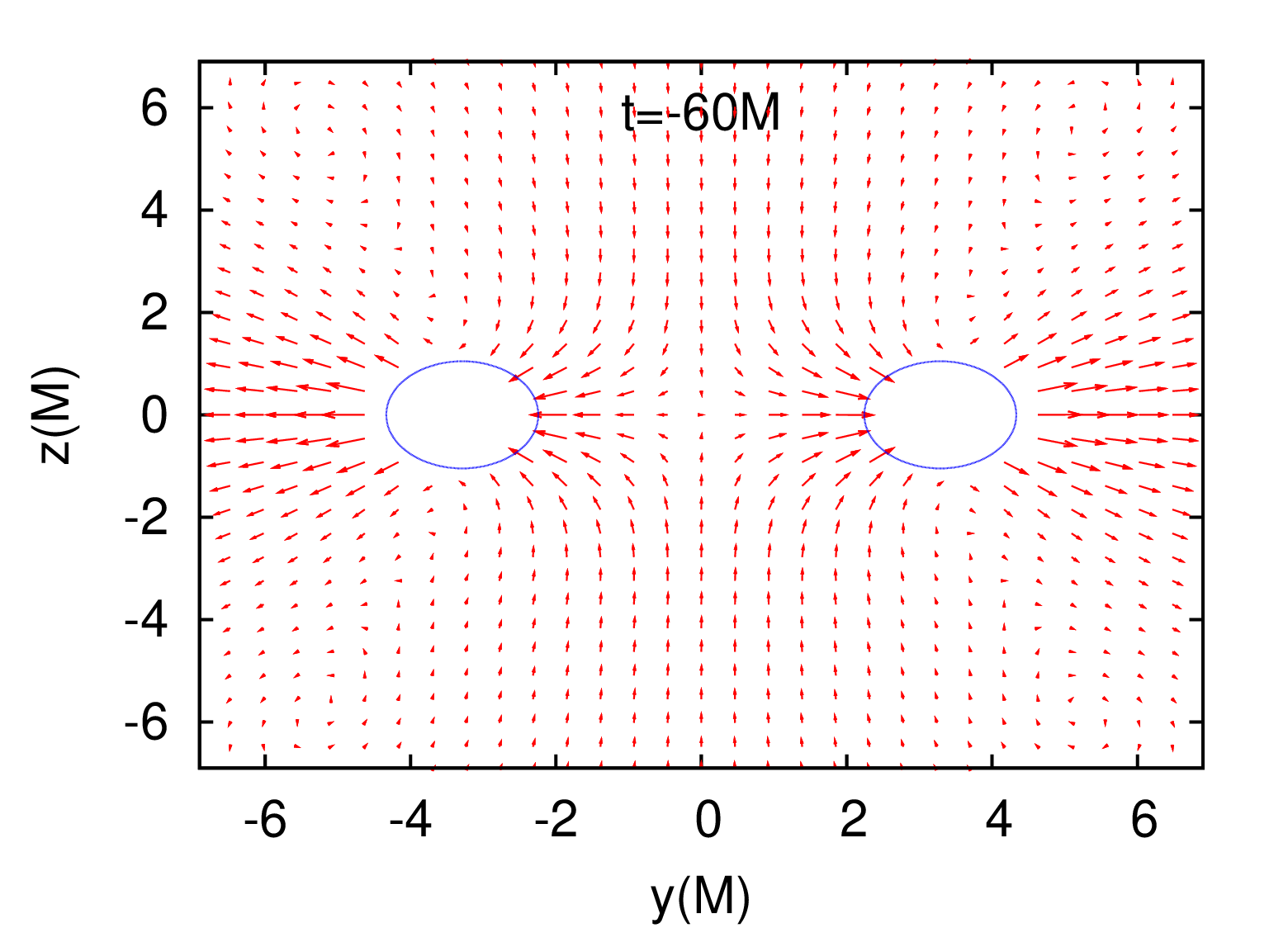,height=3.5cm,width=4.2cm} 
\epsfig{file=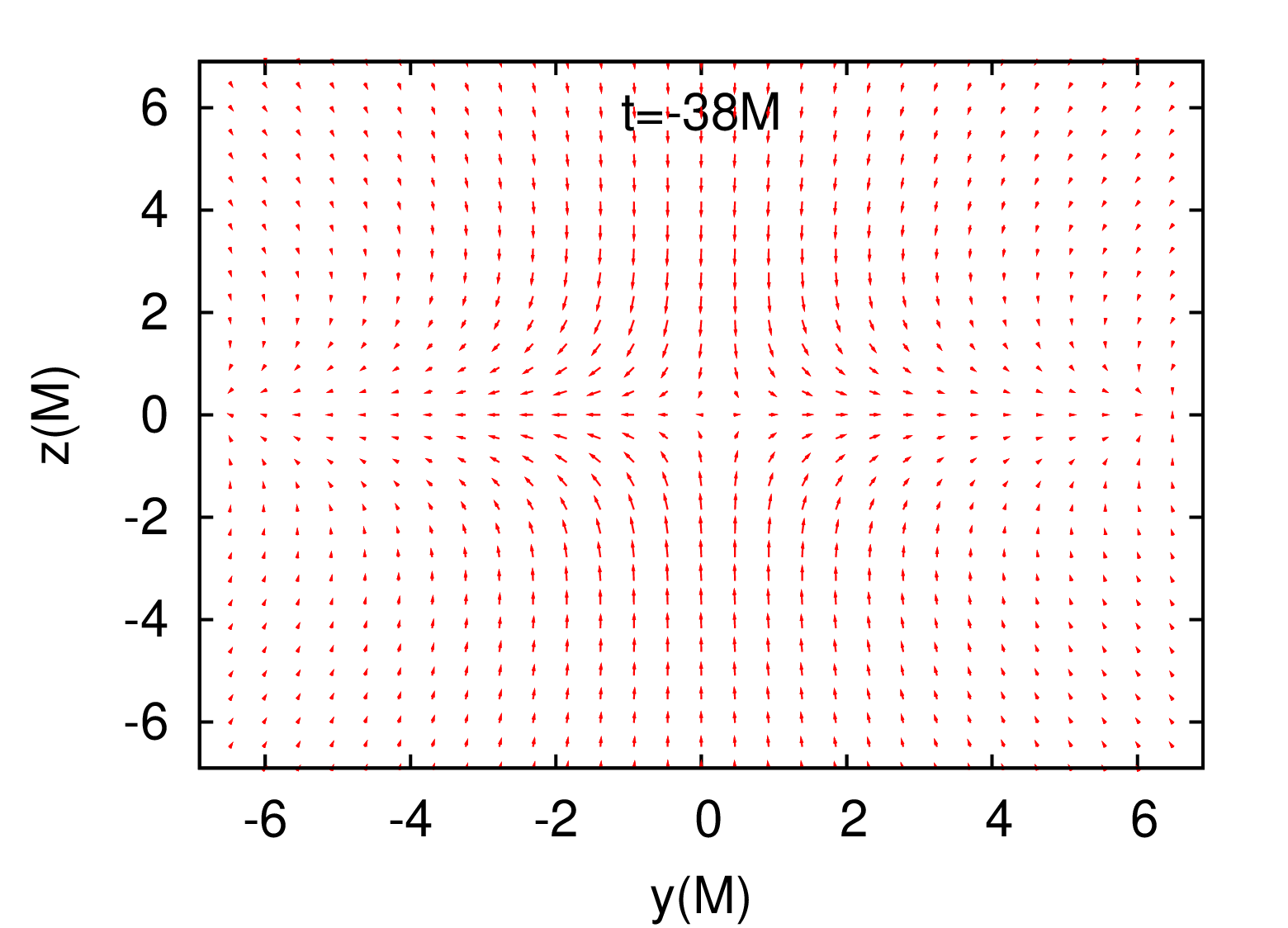,height=3.5cm,width=4.2cm} 
\epsfig{file=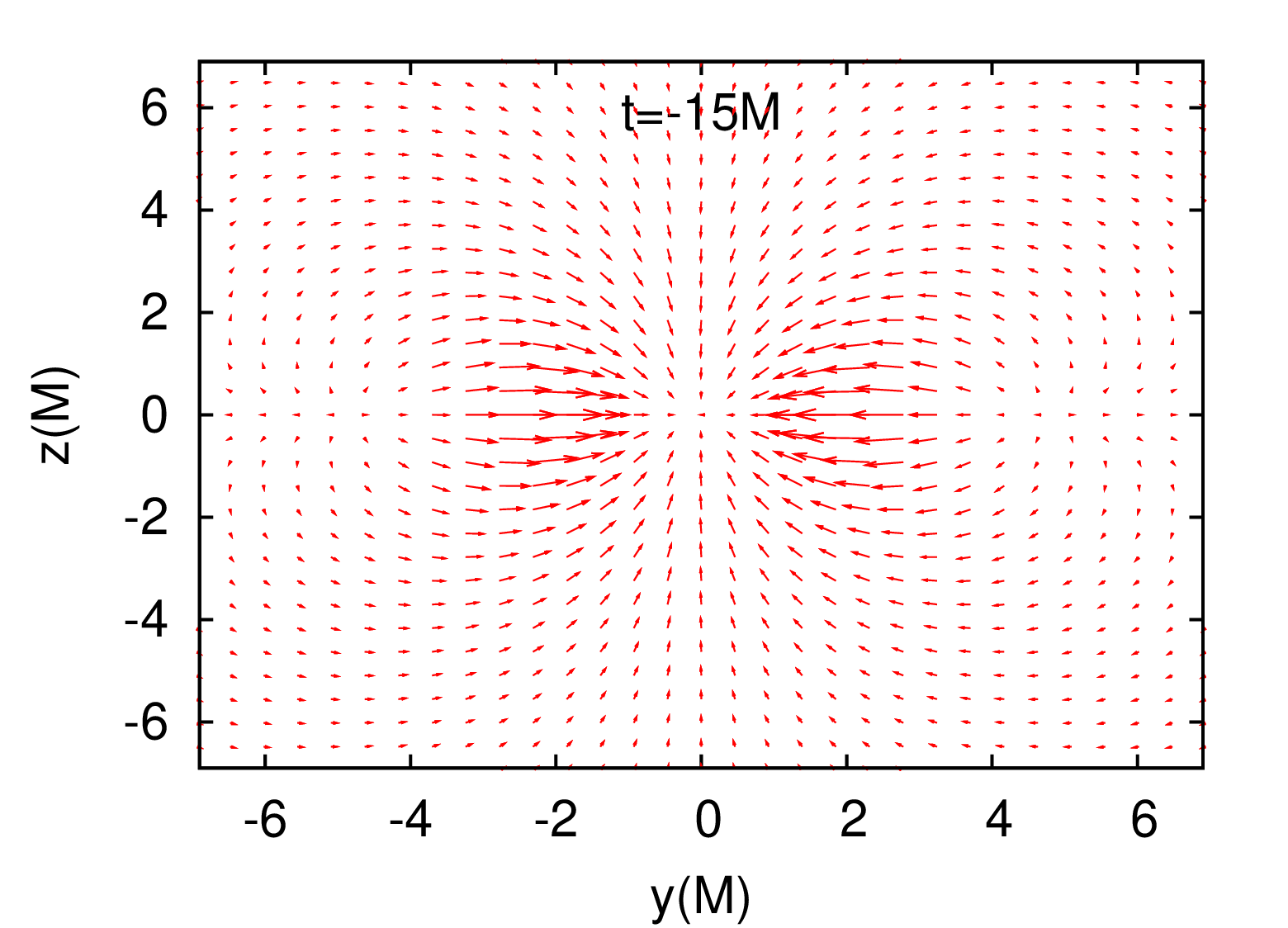,height=3.5cm,width=4.2cm} 
\epsfig{file=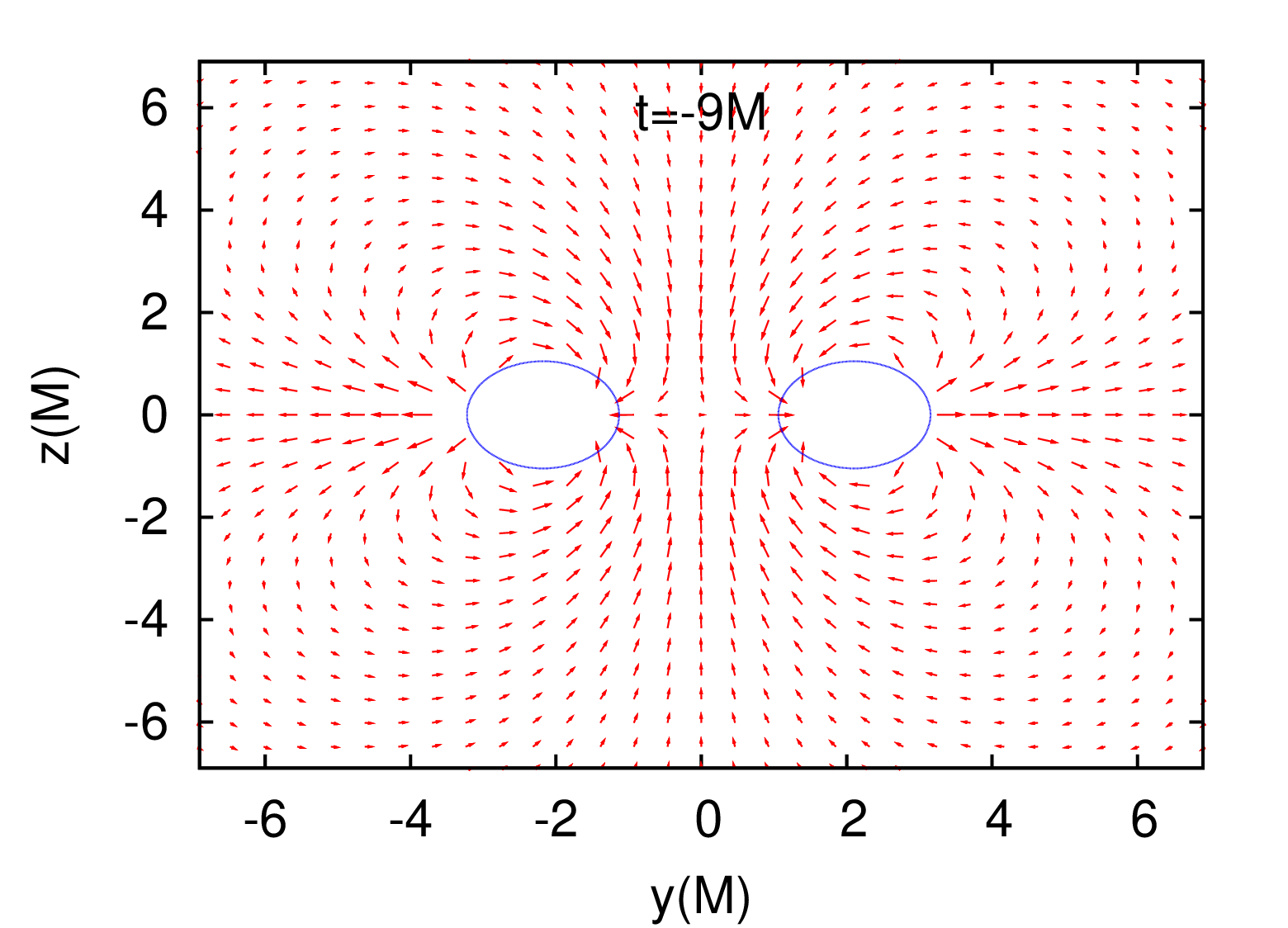,height=3.5cm,width=4.2cm} 
\epsfig{file=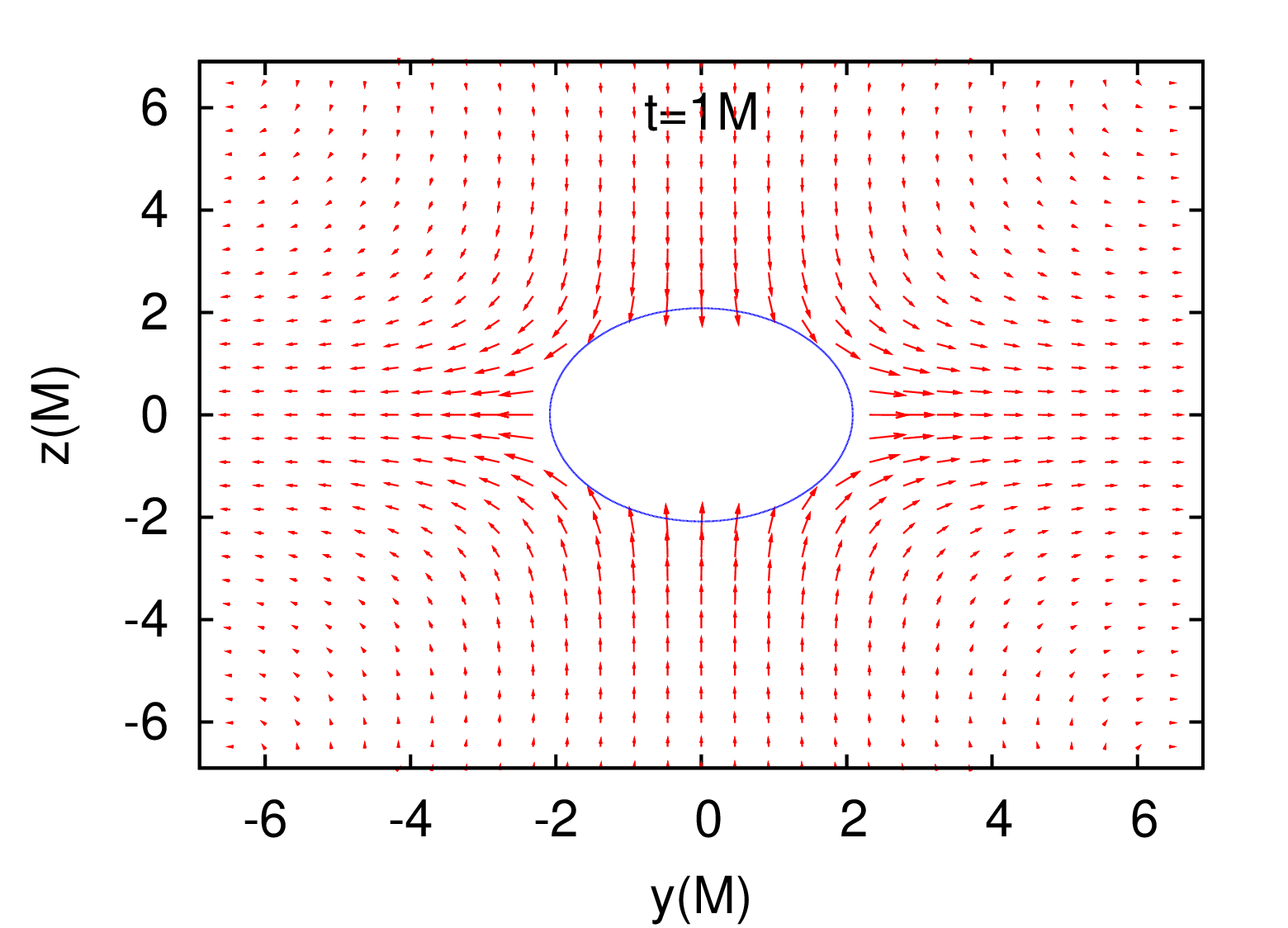,height=3.5cm,width=4.2cm} 
\epsfig{file=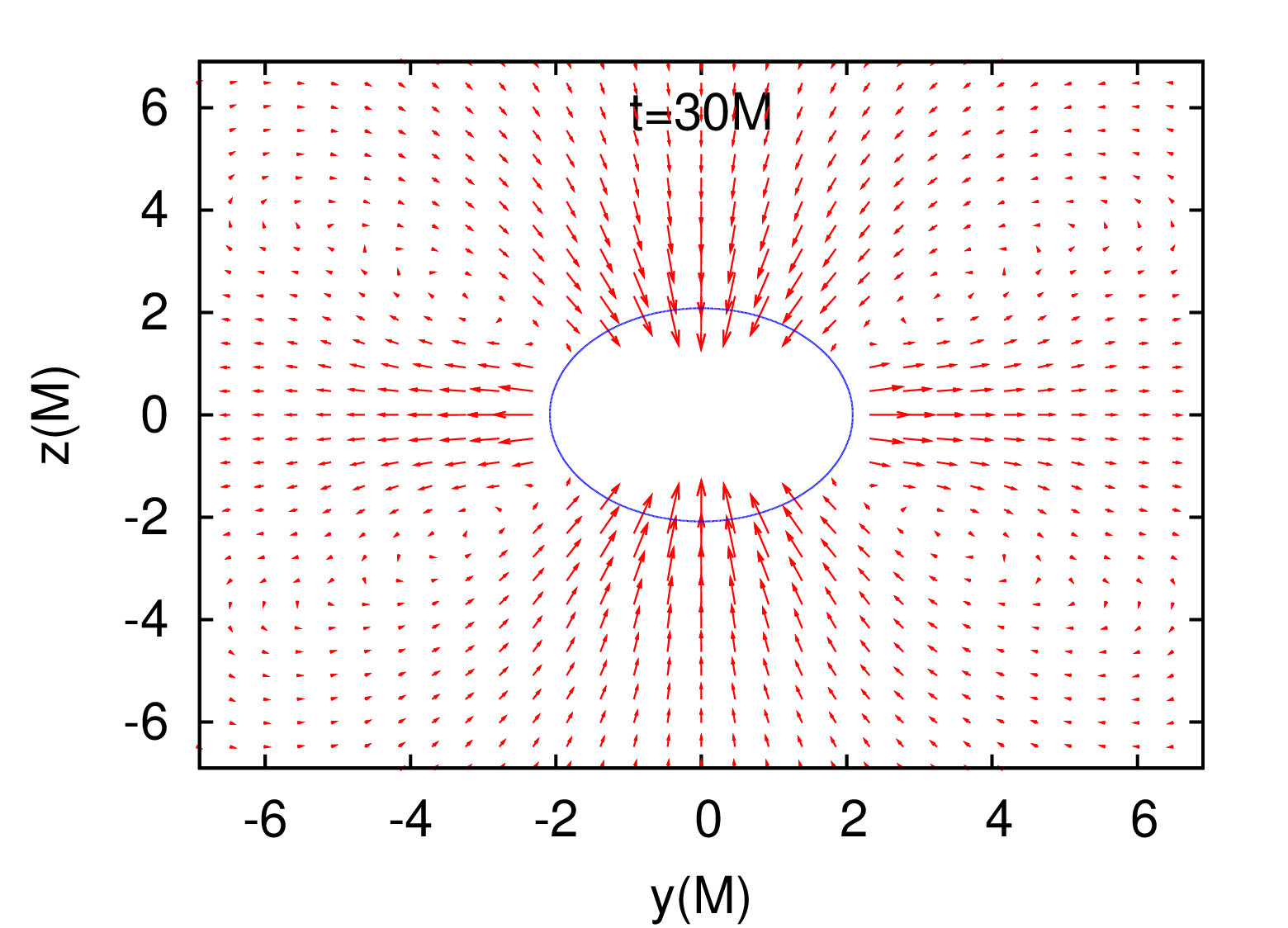,height=3.5cm,width=4.2cm} 
\caption{Electric field lines at the plane $x=0$ for the binary black hole case
at different times of the evolution. Notice that early on the field configuration is
in agreement with the expected one from the membrane paradigm with two induced
dipoles. At the black holes merge the configuration changes to a scenario
consistent with that required in the Blandford-Znajek mechanism.}
\label{fig:binary_Eyz}
\end{center}
\end{figure}

Summarizing, several distinct features in the EM fields behavior are seen during
the evolution:

$\bullet$ After an initial transient dynamics, and well before the merger takes 
place, the electromagnetic fields display a 
pattern consistent with that produced by equal dipoles orbiting about each other, 
as it is exhibited in Fig.~\ref{fig:binary_Eyz}. This can be understood from the `membrane paradigm'
point of view~\cite{1986bhmp.book.....T}, in which the horizon is endowed with a surface density of electrical
charge. The quasi-circular trajectories of the black holes cause a charge separation in the direction perpendicular to both the velocity and magnetic field, inducing an electric and magnetic field in addition to the external one
produced by the far away disk.
To understand the main features observed a simple toy model could be employed.
Such model is composed of four point charges (A,B,C,D)
that orbit circularly. Assuming perfect conductivity of the horizon,
each charge $q_i$($i=A ... D$) can be written as
\begin{equation}
 q_A = -\frac{r_H^2}{2\pi} |v\times B_o| = -q_B = q_C = -q_D,
\end{equation}
where $v$ is orbital velocity of the black holes, $B_o$ is the magnetic field
imposed by circumbinary disk and $r_{H}$ is the radius of the apparent horizon.
The orbital motion of each charge is written, in Cartesian
coordinates, as:
\begin{equation}
 \vec{r'_i}(t') = R_i(\cos\Omega t, \sin\Omega t, 0),
\end{equation}
where $R_A=R_0+r_H, ~R_B=R_0-r_H, ~R_C=-(R_0+r_H), ~R_D=-(R_0-r_H)$,
being $R_0$ the distance from either black hole to the origin (or center of mass).
The field produced by this system can be computed by 
the following procedure. First, evaluate the Lienard-Wiechert potential of
each particle and from it obtain its corresponding electric
and magnetic field contribution. Next, sum up all
contributions from the four particles to obtain the total electric and magnetic
fields. Lastly, add the externalt magnetic field to the one produced by the charges.
The final form of the fields at a given point involves implicit equations due to 
the different contributions depending on their respective retarded times. 
The intuitive picture however, can be derived easily by taking suitable limits.
For instance, when the two black holes are well separated (ie, $R_0>>r_H$)
the solution around each black hole mimics that of a dipole. At far distances
from the black holes, both the monopole and dipole contribution vanish and the system 
is described to leading order by a quadrupole. In this regime, the electric and magnetic fields at a point 
located at $||{\bf r}|| >> R_0$, have the dependence
\begin{eqnarray}\label{EB_far}
       {\bf E} \approx \frac{\Omega^3}{r} cos(2 \Omega t + \delta_E) 
 ~~,~~ {\bf B} \approx B_o + \frac{\Omega^3}{r} cos(2 \Omega t + \delta_B)
\end{eqnarray}
where $\delta_E$ and $\delta_B$ are phases whose details are not important for the present
argument. This simple form explains why $\phi_2$  and $\phi_4$ have the same frequency dependence
$2 \Omega$.

$\bullet$ Near and through the merger epoch the fields' strengths increase and so does the
flux of electromagnetic energy. Additionally the fields exhibit a configuration
consistent with that required by the Blandford-Znajek mechanism, namely a (mainly)
poloidal/toroidal magnetic/electric fields which could generate interesting energy ouput 
when interacting with a surrounding medium. As the merger approaches and take place, 
the field lines become stirred and twisted which, upon reconnection, could also release 
important amounts of energy, this scenario is beyond the scope of what we could study
with our current setup.

$\bullet$ Finally, after the merger, the system can be understood again as a conducting
sphere spinning in a external magnetic field. Its behavior can be  understood by the 
results presented in the previous section.

%
%
\section{Conclusion}

We have analyzed the behavior of electromagnetic fields influenced by
the dynamics of a binary black hole system. Our study illustrates several
interesting aspects of such systems that emit not only gravitational waves, but
can also radiate electromagnetically. Gravitational waves would be emitted
through the different dynamical stages of the system --inspiral, merger and ringdown--
through a rather smooth manner
with the peak occurring at the merger stage. Electromagnetic waves on the other hand,
will be emitted through diverse processes driven by the interaction of the EM fields
with surrounding plasma, gas or matter. In the current work we have studied the behavior
of the EM fields and illustrated their ``radiative'' behavior as energy that can
propagate outwards from the system as the black holes influence them.
This energy will likely be absorbed and re-emitted by the surrounding plasma and would,
in turn, be possible to observe. The orbiting behavior however, would leave its mark
in a time variability naturally induced in the emission process. 

Indeed, as we have illustrated here,
the EM fields have a clearly discernible pattern tied to the dynamics of the system, 
making them possible  {\em tracers of the spacetime} --in the electromagnetic sector-- 
as these features would imprint particular characteristics in processes producing observable EM signals. 
In particular, in the pre-merger stages, the black hole dynamics induce EM flux 
oscillations with a period  half that of the dominant GW signal
produced by the system, i.e., a fourth of the orbital period, and a gradual enhancement of 
the energy in the electromagnetic field. This enhancement, together with a flux of electromagnetic
energy would impact surrounding plasma in a stronger way than would be the case for a single
stationary black hole as the latter would neither exhibit an enhancement, nor would it give
rise to an outward flux of EM energy. Emissions in this later case require mechanisms like
accretion or Blandford-Znajek to take place. Certainly, the  same requirements will
apply to binary black hole systems at late times as they give rise (generically) to a single spinning black hole.
Furthermore, as the merger takes place fields are significantly twisted and stirred opening up
the possibility of interesting emissions through magnetic reconnection. A study of such scenario
requires a resistive treatment of the problem which is beyond the scope of our current work.\\

Perhaps even more exciting is the possibility of inducing a Blandford-Znajek analog for binaries as
the merger proceeds. As we have seen, the system's dynamics induces a configuration consistent
with the basic picture of this process and, by extracting rotational energy from the system,
more powerful emissions could be expected.  
For this to take place however, an important conservative requirement should hold. 
Namely, an ergosphere must be present so that rotational
energy can be extracted from the merging black holes. The best case scenario is for the ergosphere
to form before the plunging phase begins so that the black holes can orbit and sufficient
time is left for the extraction to occur.  An estimate for when this takes places
can be drawn from an ``effective one body'' approach, as described in the appendix, which
indicates highly spinning configurations are required, at least in the particle
limit, for an orbiting behavior to exist within the ergosphere. \\

Finally, and at a rather academically interesting level, it is interesting to ask what conditions
would be required to extract so much energy that the end state is essentially a non-spinning black hole
after the merger. For this to take place, 
the timescale of the BZ process ($\tau_{BZ}$) should be comparable to that of the 
merger ($\tau_M$). The latter typically
lasts $\tau_M \simeq 50 M$, the former can be estimated as in \cite{2000PhR...325...83L} giving rise
to
\begin{eqnarray}
\tau_{BZ} &\simeq& 5 \times 10^8 (10^{15} G/B)^2 (M_{\odot}/M)\,M_{\odot} , \nonumber \\
          & \simeq& 10^7 (10^{15} G/B)^2 (M_{\odot}/M)^2 \tau_M \, ; \nonumber
\end{eqnarray} 
thus, for $B > 10^{10} G$ both times result comparable for BH masses $\geq 10^{8} M_{\odot}$. 
Alternatively, one can estimate the amount of energy
that could be extracted as a function of field strength within  $\tau_M=50 M$. If $M_p$ is
the final irreducible mass of the black hole after rotational energy has been extracted via
the BZ mechanism, $M_p$ obeys
\begin{equation}
\frac{M_p}{M-M_p} \simeq 10^8 (10^{15} G/B)^2 (M_{\odot}/M)^2 \, .
\end{equation}
Thus, for $M=10^9 M_{\odot}$ and $B=\{10^7 / 10^8\} G$  
about $\{10^{-4}/ 10^{-2}\} M_{\odot} c^2$ (i.e. $\simeq \{10^{48} / 10^{50}\}$ ergs) 
is released in roughly a day.\\

While fields of these magnitudes might be unlikely, it is interesting that the 
strengths are not completely out of nature's ability to manifest.
Beyond this possibility the impact of the dynamics on the electromagnetic fields
is to induce a distinct variability as fields are
dragged by the black holes. This suggests tantalizing prospects to
detect pre-merger electromagnetic signals from systems detectable in the gravitational
wave band. However, a complete description of the problem requires the incorporation of gas and
radiation effects. Notwithstanding these missing --and important-- ingredients, 
the main qualitative features---driven by the orbiting
behavior of the black holes, whose inertia is many orders of magnitude above all 
else---would intuitively remain unaltered.

At a more speculative level, these combined signals could be 
exploited to shed light on possible observations to analyse alternative theories of gravity
where photons and gravitons might propagate at  different speeds
or gravitational energy could propagate out of our possible 4-dimensional brane
(for a recent discussion of some possibilities see~\cite{Haiman:2008zy,Bloom:2009vx}).

As a final comment we stress that while our work is a step towards understanding
possible emissions induced by binary black hole merger processes, we have only
scratched the surface of possible phenomenology. For instance, within the current approach, scenarios
with unequal masses and/or spins must be investigated and work is in progress to address
them~\cite{MPLRPY09}.  Still, a complete understanding of associated
phenomena will require investigating, in particular, the interaction with a surrounding plasma
and associated possible emission mechanisms. A preliminary related step in this direction has 
been considered in~\cite{Chang:2009rx,2008ApJ...672...83M} where the possibility of emission by a fossil
gas in between the black holes has been considered.  \\

%
%
\noindent{\bf{\em Acknowledgments:}}
We would like to thank M. Anderson,  
P. Chang, J. Frank, L. Rezzolla, S. Liebling, 
K. Menou, P. Moesta and D. Neilsen for stimulating  discussions as
well as to P. Grandclement for assistance with Lorene.
This work was supported by the NSF grants 
PHY-0803629 and PHY-0653375 and also NSERC through a Discovery Grant. 
Computations were done at TeraGrid.
LL acknowledges the Aspen Center for Physics for hospitality where
this work was started.
Research at Perimeter Institute is supported through Industry Canada
and by the Province of Ontario through the Ministry of Research \& Innovation.

\section{Appendix}
The location for the innermost stable circular orbit, at the equator,
 in a black hole spacetime 
of mass $M$ and spin parameter $a$ is given by~\cite{1972ApJ...178..347B}
\begin{eqnarray}
Z_1 &\equiv& 1 + \left(1-\frac{a^2}{M^2}\right)^{1/3}
\left[ \left(1+\frac{a}{M}\right)^{1/3} + \left(1-\frac{a}{M}\right)^{1/3} \right] \nonumber \\
Z_2 &\equiv& \left(3 \frac{a^2}{M^2} + Z_1^2 \right)^{1/2}  \nonumber \\
r_{\mbox{\scriptsize {\it ISCO}} }&=&M \left( 3+Z_2 \mp [(3-Z_1)(3+Z_1+2 Z_2)]^{1/2} \right)  \nonumber
\end{eqnarray}
The ergosphere, on the other hand, is located at

\[ r_{\mbox{\scriptsize {\it ERGO} }} = M + \sqrt{M^2-a^2 \cos(\theta)^2} \nonumber \]

\begin{figure}[h]
\begin{center}
\epsfig{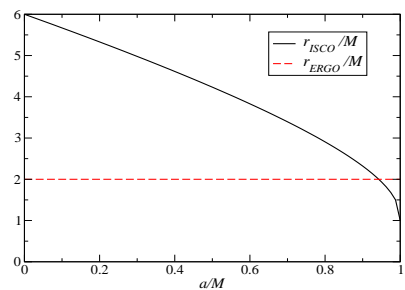}
\caption{ISCO and ergosphere radii vs $a/M$. For $a>0.943 M $ the ergosphere lies
beyond the ISCO.}
\label{fig:ISCO}
\end{center}
\end{figure}

Thus, at the equatorial plane $r_{\mbox{\scriptsize {\it ERGO} }} = 2 M$ while for prograde
orbits $r_{\mbox{\scriptsize {\it ISCO}} }(a=0)=6 M$ and $r_{\mbox{\scriptsize {\it ISCO}} }(a=M)= M$. Consequently, for
sufficiently high spins $r_{\mbox{\scriptsize {\it ERGO} }} > r_{\mbox{\scriptsize {\it ISCO}} }$. This is illustrated
in figure~\ref{fig:ISCO}, where the critical value at which the two lines cross
A related interesting point is that this argument bears relevance also to
the generation of gravitational waves themselves. If the black holes
orbit outside the isco but inside the ergosphere they could tap rotational
energy and produced stronger emissions. All cases so far studied numerically lie
below this critical value and so, if this simplistic model holds, it would
indicate simulations  have not yet probed the possibility of extraction
of rotational energy. Therefore binary black hole systems could still potentially yield 
further interesting features in such regime.

%
%
\bibliography{./bhbhem}
\bibliographystyle{apsrev}

%
%
\end{document}